\documentclass[graphicx,superscriptaddress,longbibliography,prl,reprint]{revtex4-1}

\usepackage[T1]{fontenc}
\usepackage{siunitx}
\usepackage{graphicx}
\usepackage[table,xcdraw]{xcolor}
\usepackage[utf8]{inputenc}

\usepackage[normalem]{ulem}
\useunder{\uline}{\ul}{}

\begin{document}

\title{Mass measurement of graphene using quartz crystal microbalances}

\author{Robin J. Dolleman}
\affiliation{Kavli Institute of Nanoscience, Delft University of Technology, Lorentzweg 1, 2628 CJ, Delft, The Netherlands}
\affiliation{Current affiliation: 2nd Institute of Physics, RWTH Aachen University, 52047 Aachen, Germany}
\email{dolleman@physik.rwth-aachen.de}
\author{ Mick Hsu}
\affiliation{Kavli Institute of Nanoscience, Delft University of Technology, Lorentzweg 1, 2628 CJ, Delft, The Netherlands}
 \author{Sten Vollebregt}
 \affiliation{Department of Microelectronics, Delft University of Technology, Feldmannweg 17, 2628CT, Delft, The Netherlands}
 \author{John E. Sader}
\affiliation{ARC Centre of Excellence in Exciton Science, School of Mathematics and Statistics, The University of Melbourne, Victoria 3010, Australia}
\author{ Herre S. J. van der Zant}
\affiliation{Kavli Institute of Nanoscience, Delft University of Technology, Lorentzweg 1, 2628 CJ, Delft, The Netherlands}
\author{ Peter G. Steeneken}
\affiliation{Kavli Institute of Nanoscience, Delft University of Technology, Lorentzweg 1, 2628 CJ, Delft, The Netherlands}
\affiliation{Department of Precision and Microsystems Engineering, Delft University of Technology, Mekelweg 2, 2628 CD, Delft, The Netherlands}
\author{ Murali K. Ghatkesar}
\affiliation{Department of Precision and Microsystems Engineering, Delft University of Technology, Mekelweg 2, 2628 CD, Delft, The Netherlands}
\email{M.K.Ghatkesar@tudelft.nl}

\begin{abstract}
Current wafer-scale fabrication methods for graphene-based electronics and sensors involve the transfer of single-layer graphene by a support polymer. This often leaves some polymer residue on the graphene, which can strongly impact its electronic, thermal, and mechanical resonance properties. To assess the cleanliness of graphene fabrication methods, it is thus of considerable interest to quantify the amount of contamination on top of the graphene. Here, we present a methodology for direct measurement of the mass of the graphene sheet using quartz crystal microbalances (QCM). By monitoring the QCM resonance frequency during removal of graphene in an oxygen plasma, the total mass of the graphene and contamination is determined with sub-graphene-monolayer accuracy. Since the etch-rate of the contamination is higher than that of graphene, quantitative measurements of the mass of contaminants below, on top, and between graphene layers are obtained. We find that  polymer-based dry transfer methods can increase the mass of a graphene sheet by a factor of 10. The presented mass measurement method is conceptually straightforward to interpret and can be used for standardized testing of graphene transfer procedures in order to improve the quality of graphene devices in future applications. 
\end{abstract}

\maketitle

\section{Introduction}
The remarkable electronic \cite{geim2007rise,neto2009electronic}, thermal \cite{balandin2008superior,pop2012thermal,dolleman2017optomechanics} and mechanical \cite{lee2008measurement,davidovikj2017nonlinear,sajadi2017experimental} properties of graphene have opened the door for many new electronic devices \cite{song2011stamp,chen2013graphene,lemme2007graphene,gimod} and sensors \cite{dauber2015ultra,smith2017graphene,ricciardella2017effects,smith2013pressure,dolleman2015graphene,dolleman2016graphene,vollebregt2017suspended,davidovikj2017static}. Fabrication of these devices on wafer-scale often requires transfer of sheets of single-layer graphene grown by chemical vapor deposition, using a support polymer \cite{suk2011transfer,li2009large,li2009transfer,lee2010wafer,chen2013high,zande2010large}. It is inevitable that this introduces some transfer contamination on top of the graphene \cite{her2013graphene,lin2011graphene}, significantly impacting the device's electronic \cite{pirkle2011effect,suk2013enhancement,chan2012reducing,goossens2012mechanical,moser2007current}, thermal \cite{pettes2011influence,jo2015reexamination}, or mechanical resonance properties \cite{bunch2008impermeable,singh2010probing,barton2012photothermal,song2011stamp,chen2009performance}. A simple and accurate test to determine the amount of contamination on top of graphene is therefore of high interest to the community. However, due to the optical transparency, softness and small thickness of the contamination layers, with current popular characterization techniques such as Raman spectroscopy \cite{ferrari2006raman}, optical microscopy \cite{blake2007making} and atomic force microscopy \cite{nemes2008anomalies} it is difficult to detect, and even more difficult to quantify, the amount of contamination on top of large sheets of graphene. 

Several works have determined the amount of contamination on top of graphene resonators by tracking the resonance frequency shift in response to an out-of-plane force \cite{bunch2008impermeable,singh2010probing,barton2012photothermal,song2011stamp,chen2009performance}. However, these methods require knowledge of the mechanical properties of the resonator, which vary considerably from device-to-device, impacting the accuracy of resonance-based measurement methods \cite{davidovikj2017nonlinear,nicholl2015effect,lee2012estimation,ruiz2011softened,davidovikj2016visualizing,wong2010characterization}. 
Moreover, resonance based methods only probe contamination over a small area of the suspended resonator, whereas large lateral variations in the amount of contamination can occur. For assesment of production techniques, it is important to have procedures that ensure low contamination levels over large areas. 

Here, we present a method to determine the mass of graphene and of the contamination layers on top of graphene, between graphene double-layers, and below graphene. We employ quartz crystal microbalances (QCM), which are piezoelectric quartz crystals that can be brought into resonance by applying an oscillating voltage \cite{Sauerbrey1959}. QCMs are popular tools to measure growth rates during thin film deposition and in biochemical applications \cite{o1999commercial,buttry1992measurement}, because of their simplicity and high accuracy. In this work we demonstrate the use of QCMs to determine the mass of graphene and contaminants by an in-situ measurement during oxygen plasma etching. In contrast to mechanical resonance based methods, the proposed method is no longer sensitive to the mechanical properties of the graphene and thus facilitates a direct measurement of the mass and furthermore allows large areas of graphene to be studied. 

\begin{figure*}[t!]
\includegraphics{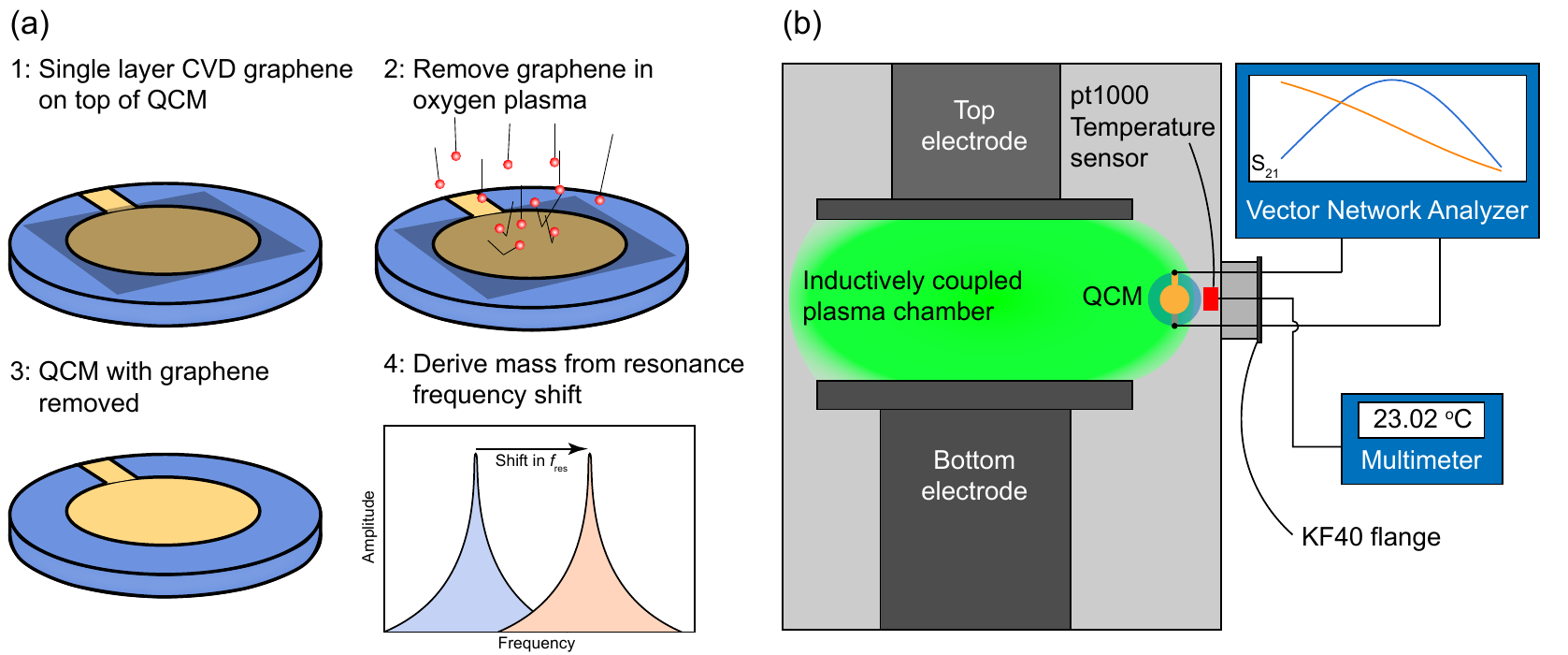}
\caption{(a) Experimental procedure to measure the mass of graphene using the quartz crystal microbalance (QCM). The measurement starts with a QCM with a sheet of CVD graphene covering one of the electrodes. The oxygen plasma etches away the graphene and any contaminants until the etching stops. Continuous monitoring of the resonance frequency of the crystal allows one to determine the mass that has been removed by the plasma. (b)  Experimental setup to determine the mass of graphene during plasma etching. The QCM and temperature sensor are mounted on a KF40 flange with electrical feedthroughs to an ICP etching chamber. Outside the vacuum chamber the oscillator circuit processes and conditions the signal from the QCM which is then read-out using a frequency counter. A platinum resistor pt1000 thermometer determines the temperature near the QCM. \label{fig:QCMprocedure}}
\end{figure*}
\section{Experimental setup and procedure}\label{sec:exp}
The sensors consist of AT-cut piezoelectric quartz crystals (Noveatech S.r.l. AT10-14-6-UP) between two gold contacts, vibrating at a resonance frequency near 10 MHz. A Piranha solution and oxygen plasma treatment on both sides are used to clean the sensors of all organic contaminants, no significant organic contamination remains after this process as shown in Supporting Information S3. Large sheets of graphene grown by chemical vapour deposition (CVD) are transferred on top of one of the electrodes using a widely used dry transfer method \cite{suk2011transfer}. It is ensured that the graphene sheet fully covers the electrode. A second layer of graphene is transferred on some of the crystals to create double-layer graphene. No attempts are performed to clean the graphene after transfer, since we are interested to see the amount of transfer residue on top of the graphene as a result of this process. 

Figure \ref{fig:QCMprocedure}(a) shows the experimental protocol to measure the mass of graphene. The crystal with CVD graphene is placed in the plasma chamber. Oxygen plasma etches the graphene and organic residues, which reduces the mass on top of the crystal that results in an increase of the QCM resonance frequency. The resonance frequency of the QCM is continuously monitored during the etching process. Stabilization of the resonance frequency indicates full removal of the graphene and all the organic contaminants from the QCM. The shift in the resonance frequency $\Delta f$ can be related to the removed mass using the Sauerbrey equation \cite{Sauerbrey1959}:
\begin{equation}
\Delta f = - C \rho h,
\end{equation}
where $\rho$ is the density, $h$ the thickness of the material on top of the QCM, and the constant $C$ is given by the properties of the quartz crystal:
\begin{equation}
C= \frac{2}{\sqrt{\rho_q \mu_q}} f_0^2,
\end{equation}
where $\rho_q$ is the density of quartz, $\mu_q$ the shear modulus and $f_0 = 10$ MHz the unloaded resonance frequency of the crystal. For the crystals used in this work, a single monolayer of graphene with a mass density of $\rho h_{\mathrm{graphene}} = 0.76$ mg/m$^2$ corresponds to a theoretical resonance frequency shift of 17.19 Hz.

Figure \ref{fig:QCMprocedure}(b) shows the experimental setup to measure the resonance frequency of the crystal during etching. We use a reactive ion etcher (Leybold Hereaus Fluor F2) in a class 10000 (ISO7) cleanroom as the plasma chamber. A blind vacuum flange is adapted to create electrical feedthroughs to the chamber, and connected to a KF-40 port on the plasma chamber. A vector network analyzer (VNA) interrogates the resonance frequency of the membrane by a transmission measurement. Alternatively one can use a commercially available oscillator circuit as shown in the Supporting Information S1. However, the VNA produces the best results as it is not sensitive to the interference from the RF plasma at 13.56 MHz-- the homodyne detection scheme rejects these frequencies. The oscillator circuit, on the other hand, will often lose lock on the mechanical resonance frequency when the plasma ignites, eliminating measurement data during the etching process (see Supporting Information S3). A pt-1000 temperature sensor is placed in the chamber near the crystal to monitor the temperature, because this affects the resonance frequency. The uncertainty in the resonance frequency is determined by three factors, as shown in the Supporting Information S2. First, the frequency dependence of the crystal on temperature, which is characetrized in a subsequent measurement. Second, the occurrence of a small ($\sim$3 Hz) jump when the RF power is switched on, and third, the occurrence of random frequency jumps during the measurements which possibly occur due to spurious modes. 

\section{Results}\label{sec:results}
\begin{figure*}[t!]
\includegraphics{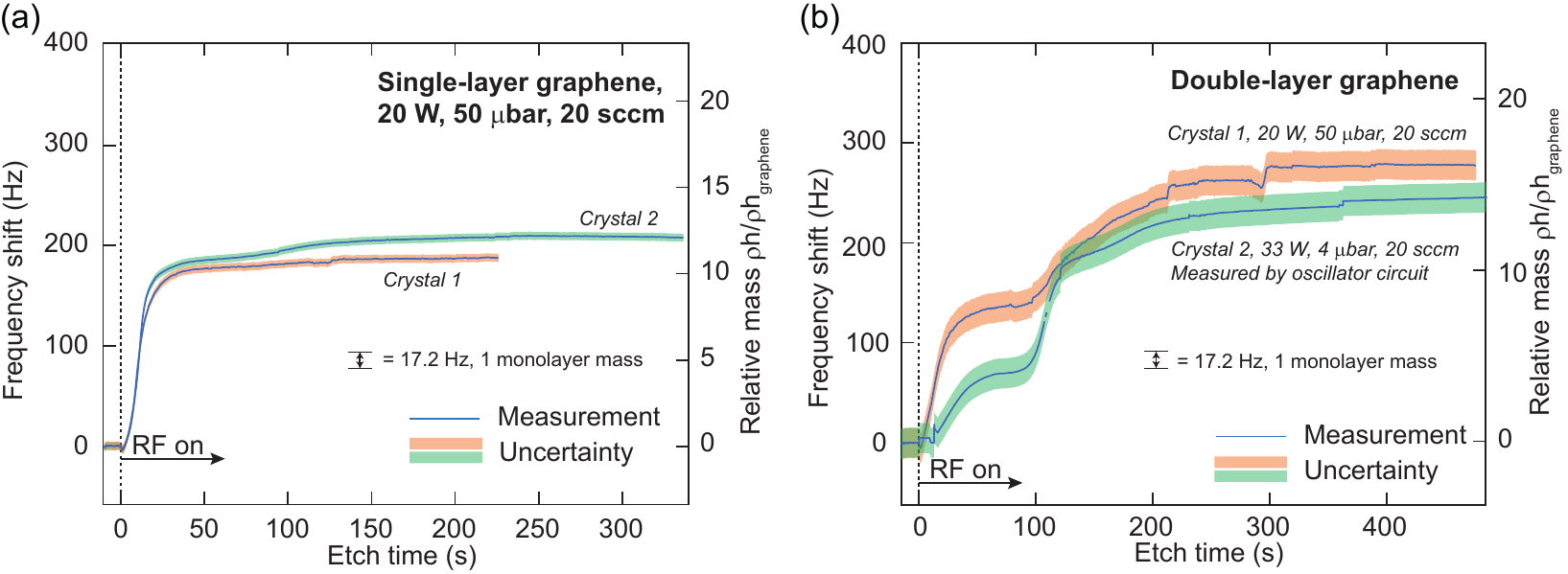}
\caption{ (a) Frequency shift as a function of etch time for two crystals covered by single layer graphene. At time 0 the oxygen plasma is switched on. Both crystals show an initial high etch rate which rapidly decreases after $\sim$20 seconds, followed by a slight increase until the frequency settles. The uncertainty bar is determined by (1) the temperature dependence of the frequency, (2) a small jump in frequency when the RF power is switched on and (3) the occurrence of sudden jumps in the frequency (see supplemental information). The total uncertainty is determined to be 7 Hz.  (b) Frequency shift for two crystals covered with double-layer graphene. These crystals show a striking decrease in etching rate, after which the etching rapidly increases again. At approximately 130 seconds, crystal two features a second decrease in the etch rate. The uncertainty is larger with respect to the single-layer devices due to the more frequent occurrence of jumps in the frequency.  \label{fig:QCMresults}}
\end{figure*}
Figure \ref{fig:QCMresults} shows the resonance frequency of four graphene-covered sensors as a function of the etch time in an oxygen plasma. Samples covered with single-layer graphene are shown in Fig. \ref{fig:QCMresults}(a) . After the plasma is switched on, the resonance frequency rapidly increases which indicates the removal of mass from the sensor. We observe that after approximately 20 seconds, the etch rate decreases considerably. After some time, however, the etch rate slightly increases again before the frequency stabilizes. In crystal 1, this corresponds roughly to a monolayer of graphene being etched, while in crystal 2 more mass is removed in this slower regime. A total mass corresponding to $10.8\pm0.4$ layers of graphene is removed from crystal 1, while a total of $12.2\pm0.4$ layers was removed from crystal 2.  The total uncertainty of these measurements is determined to be 7 Hz, thus achieving sub-monolayer accuracy in the mass. 

Results on the stacked double layers of graphene are shown in Fig. \ref{fig:QCMresults}(b). Similar to the case of the single-layer graphene, the etching slows down considerably after approximately 40 to 50 seconds. After this, the etching rate increases significantly. In crystal 2, a second decrease in the etching rate is observed at approximately 130 seconds of etching time. Later, the etching rate increases again and finally stops. In crystal 1, this second decrease in etching rate is less prominent due to spurious frequency jumps near 200 seconds and 300 seconds. A total mass corresponding to $16.0 \pm 1.1$ monolayers of graphene is removed from crystal 1, while $14.0 \pm 1.1$ monolayers are removed from crystal 2. The more frequent occurrence of random jumps during the measurements of these samples cause a higher uncertainty compared to the single-layer crystals. Possible causes of the random jumps are outlined in the Supporting Information S2.  

The total etched mass of all the sensors used in this work are shown in Table \ref{tab:results}. The remaining etching curves that are not shown in Fig. \ref{fig:QCMresults} are given in the Supporting Information S3. For the double layer crystals, we etch away on average an equivalent mass of 16.7 monolayers of graphene while for the single layers we etch away 11.8 layers of graphene. 

\begin{table}[]
\caption{Measured mass per unit square $\rho h$ divided by the theoretical mass of graphene $\rho h_{\mathrm{graphene}} = 0.76$ mg/m$^2$. The SLG column shows single layer graphene samples and the DLG column double layer graphene. Three samples ($^{\dagger}$) were measured by the VNA and the remainder ($^{\ddagger}$) were measured by the oscillator circuit.  \label{tab:results}}
\begin{tabular}{|l|l|l|}
\hline
Sample & SLG $\rho h/ \rho h_{\mathrm{graphene}}  $                     & DLG $\rho h/ \rho h_{\mathrm{graphene}}  $               \\ \hline
1      & {12.2}$^{\dagger}$ & {16.0}$^{\dagger}$ \\ \hline
2      & {10.8}$^{\dagger}$ & {14.0}$^{\ddagger}$ \\ \hline
3      & {10.1}$^{\ddagger}$ & {19.7}$^{\ddagger}$ \\ \hline
4      & {14.0}$^{\ddagger}$ & {19.2}$^{\ddagger}$ \\ \hline
5      & { }     & {14.5}$^{\ddagger}$ \\ \hline
\end{tabular}
\end{table}

\section{Discussion}\label{sec:disc}
We first discuss the observed variation in the etch rate, in particular the slow etching regimes in the double-layer samples. By comparing the single layer and double layer samples in Fig. \ref{fig:QCMresults}, we conclude that the slower etch rate can be attributed to the graphene, since the double layer samples typically show a second decrease in the etching rate. The results thus show that graphene etching rate is much slower than the contamination. From the data with SLG in Fig. \ref{fig:QCMresults}(a), we conclude that most of the contamination is present on the top side of the graphene since initially most of the mass is removed and after the etching slows down it does not increase again. This makes transfer residues from the polymer a likely source of the contamination. On Crystal 2 with SLG, significantly more mass (approximately 1.5 graphene monolayers) is removed in the slow etching regime, which might indicate the presence of contamination underneath the graphene. Possible sources of contamination underneath graphene could be water that is known to accumulate between graphene and the substrate \cite{doi:10.1021/ar500306w}, or insufficient cleaning of organic contaminants from the QCM before transfer. In the Supporting Information S3, we show a measurement at much lower plasma power of 4W. At these powers, graphene was not fully removed but rather oxidized as shown by Raman spectroscopy, confirming the slow etching rate of graphene with regard to the contamination. 

The measurements of double layer graphene samples in Fig. \ref{fig:QCMresults}(b) provide further evidence that the slower etch rate can be attributed to the graphene.  One interpretation of the data is that the etching of the first graphene layer causes a significant decrease in the etching rate after 40 seconds and the less-prominent second decrease can be attributed to the second graphene layer. Since the etching rate significantly increases after slowing down, we conclude that the stacking of these layers effectively results (from bottom to top) in a graphene-residue-graphene-residue stack. While the first region with a slow etch rate shows a relatively sharp decrease in etching rate, the second slow regime is considerably smoother. This may be due to lateral non-uniformities in the etch rate or contamination thickness, which result in variations in the time when the second graphene layer is reached and etched. One striking observation is that the addition of the second graphene layer increases the mass by only $\sim$40\%, instead of doubling the mass. From the experiments (Fig. \ref{fig:QCMresults}) it appears that the additional mass can be attributed to the top contamination layer of the single layer graphene, while this layer is significantly thinner in the double-layer samples. The underlying cause of this observation is currently unknown. 

The method presented here can be used in several future technological applications. For example, the frequency range of tunable oscillators \cite{chen2013graphene,davidovikj2017thermal,ye2017very} and the responsivity of resonating pressure sensors \cite{bunch2008impermeable,dolleman2015graphene,dolleman2016graphene,vollebregt2017suspended} is significantly impacted by the mass-per-unit-square of the device. The method is also a useful technique to determine the presence of contamination. The electron mobility of graphene, for example, is significantly impacted by contaminants \cite{pirkle2011effect,suk2013enhancement,chan2012reducing,goossens2012mechanical,moser2007current}. Graphene has also been proposed as heat spreaders for thermal management in electronic circuits \cite{ghosh2008extremely}, but contaminated samples also show a significantly lower thermal conductivity \cite{pettes2011influence,jo2015reexamination}. For upscaling electronic and thermal graphene applications to the wafer scale, the proposed QCM method can be used to select the best transfer technique to produce high-quality graphene devices. Furthermore, the QCM method no longer requires the fabrication and testing of devices to optimize the transfer procedure \cite{pirkle2011effect,suk2013enhancement,chan2012reducing,goossens2012mechanical,moser2007current}, which simplifies the procedure and improves the throughput of the optimization process. The QCM method thus enables large-scale quality control of graphene sheets. Moreover, since graphene etches much slower than the contamination, the technique can discriminate between the amount of contamination underneath and on top of the graphene. 

For research, the method is useful to study the mass of wrinkled graphene membranes to reveal their hidden area \cite{PhysRevLett.118.266101}, for example graphene transferred on smooth or rough substrates. Furthermore, the QCM measurement can be a corroboration to the number of layers revealed by Raman spectroscopy \cite{ferrari2006raman}. In particular, the method is accurate enough to count the number of layers on few-layer graphene samples, in the regime where this is difficult to achieve with Raman spectroscopy. Moreover in heterostructures or other stacks of multilayer 2D materials the QCMs are useful to reveal the presence of trapped residual materials, which can hamper the interlayer coupling that gives these stacks their favourable properties.

Future work on this QCM method can focus on high-temperature measurements, enabling one to study and optimize annealing proceduces to clean graphene \cite{pirkle2011effect,lin2011graphene,cartamil2017mechanical}.  Recent advances in microbalance technology has resulted in GaPO$_4$ crystals with operating temperatures exceeding 900$^{\circ}$C, which could even facilitate in-situ study of chemical vapour deposition of graphene on certain metal thin-films  \cite{thanner2002gapo4,Thanner2003}. Furthermore, the applicability of QCM's in liquids could also be useful to study chemical cleaning of graphene \cite{her2013graphene,liang2011toward,suk2013enhancement} in-situ and optimize the process. This would in particular be interesting with regard to oxide and metal contaminants which are not removed by oxygen plasma. It would also be interesting to compare and benchmark transfer methods, for example transfer in air versus vacuum \cite{LEE2015286} or wet versus dry transfer techniques. 

\section{Conclusion}\label{sec:conc}
We present a method to determine the mass-per-unit-square and etch rate of CVD-grown graphene sheets using quartz crystal microbalances. This is achieved by etching graphene on a QCM in oxygen plasma and measuring the resonance frequency of the crystal in-situ. We find that by using a widely used dry transfer method, the mass of single-layer graphene sheet is observed to be ten times higher than the theoretical mass of graphene. The time-dependence of the etching rate shows that most of the contamination is on top of the graphene. The method is useful for quality control of large sheets of graphene for future sensing, electronic and thermal applications. 

\section*{Acknowledgements}
\begin{acknowledgements}
The authors thank Applied Nanolayers B.V. for supply and transfer of the single-layer graphene and Hugo Solera Licona for help with cleaning the crystals. This work is part of the research programme Integrated Graphene Pressure Sensors (IGPS) with project number 13307 which is financed by the Netherlands Organisation for Scientific Research (NWO).
The research leading to these results also received funding from the European Union's Horizon 2020 research and innovation programme under grant agreement No 785219 Graphene Flagship. The authors acknowledge support from the Australian Research Council Centre of Excellence in Exciton Science (CE170100026) and the Australian Research Council Grants Scheme.
\end{acknowledgements}

\newpage
\onecolumngrid
\appendix
\newpage

\section*{Supporting Information}
\section*{S1: Commercially available oscillator circuit}
\begin{figure}[h!]
\centering
\includegraphics{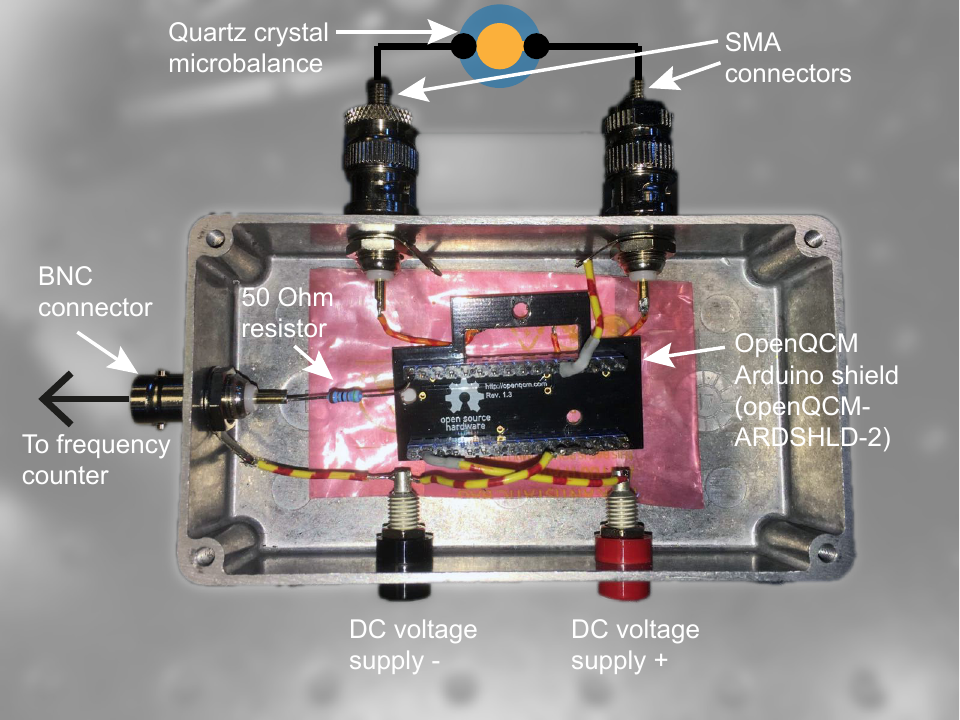}
\caption{Photograph of the metal boxes containing the openQCM oscillator circuit. \label{fig:openQCM}}
\end{figure}
Part of the measurements in this work are performed by using a commercially available oscillator circuit mounted in a metal box as shown in Fig. \ref{fig:openQCM}. This is the openQCM Arduino shield (openQCM-ARDSHLD-2), where the crystal is normally mounted directly to the QCM using a holder. However, our application of the crystal in vacuum requires the PCB to be mounted outside the chamber. Therefore, the PCB is placed in a metal box and the connections, that normally connect directly to the crystal, are now connected to the center pins of two SMA connectors. SMA cables can then be connected to the KF-40 vacuum flange, which also uses SMA connections for the electrical feedthroughs. The DC voltage, that is normally supplied by an Arduino, is now supplied by an external voltage source (Rigol DP832A). A BNC connector carries the signal from the PCB to the frequency counter (Keysight 53230A) through a 50 Ohm resistor. This resistor matches to the load of the frequency counter and it is found that this drastically improves signal quality. 

\section*{S2: Determination of uncertainty in the frequency shift}
\begin{figure}[h!]
\centering
\includegraphics{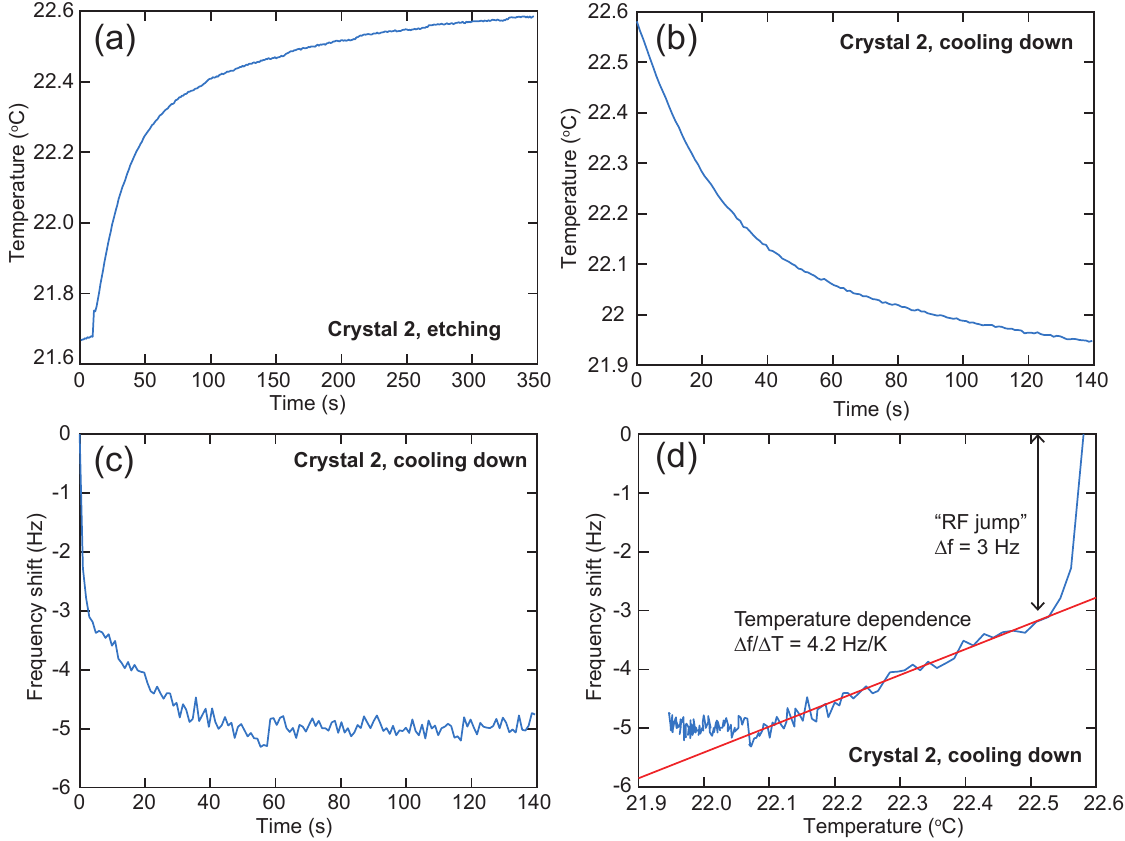}
\caption{Determination of the uncertainty of the frequency shift for crystal 2. (a) Temperature as function of time during the etching, the corresponding frequency shift is shown in Fig. 2(a) in the main text. (b) Temperature as function of time after etching, the RF power is switched off at $t = 0$ s. (c) Frequency shift as function of time after etching. (d) Frequency as function of temperature during cooling down, this graph is used to determine the temperature sensitivity of the crystal. \label{fig:uncertainty}}
\end{figure}
The quartz crystal microbalance is a highly sensitive platform for mass measurement, but is also sensitive to temperature and other effects which leads to some uncertainty in the measurements. For this reason, the temperature in the chamber is monitored during the etching as shown in Fig. \ref{fig:uncertainty}(a). To give an estimate of the uncertainty, after each etching procedure the measurement of the frequency and temperature continued after the RF power was switched off. During this measurement the temperature drops (Fig. \ref{fig:uncertainty}(b)) and a frequency shift is observed (Fig. \ref{fig:uncertainty}(c)). Also, immediately after switching on the RF power a $\sim$3 Hz jump in the frequency is observed. The cause of this is not known, therefore we add this to the uncertainty as well. By plotting the frequency shift versus the temperature in Fig. \ref{fig:uncertainty}(d), we determine the sensitivity of the resonance frequency to the temperature. In this crystal, the total temperature change during etching is 0.95K, resulting in a total temperature uncertainty of 4 Hz. The total uncertainty including the RF jump is then determined to be 7 Hz. In all crystals we found similar values. 

  For the double layer crystals the uncertainty is higher due to the occurrence of spontaneous jumps in the frequency. Most likely these can be attributed to spurious modes in the crystal or coupling to other modes. However, we cannot rule out that these jumps occur due to particles attaching or detaching to the crystal, therefore the height of these jumps was included in the uncertainty. 

\section*{S3: Additional measurements}
\begin{figure}[h!]
\centering
\includegraphics{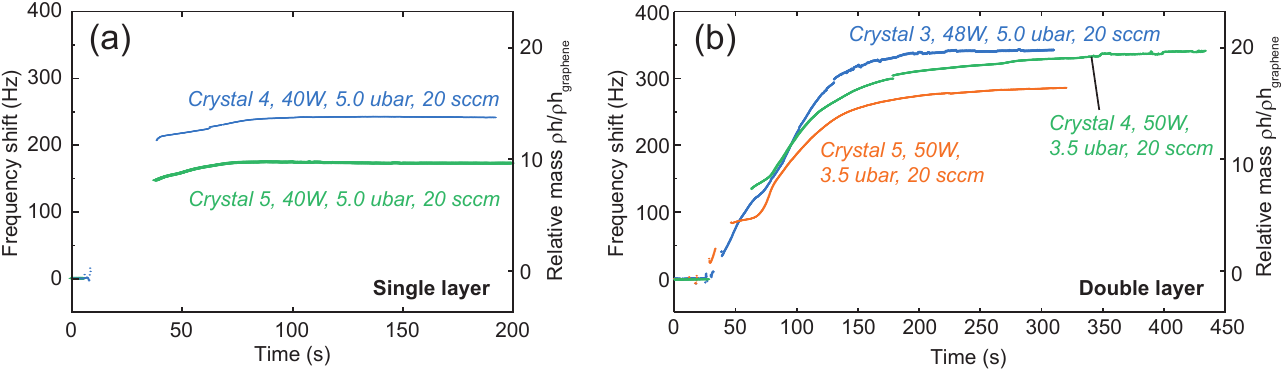}
\caption{Additional frequencies as function of time measured by the oscillator circuit. Note that most curves are missing data due to the oscillator circuit becoming unstable once the RF power is switched on, only the start and end frequencies are used to determine the etched mass in Table 1 in the main section of the paper. (a) Two measurements on crystals with single layer graphene. (b) Three crystals with double-layer graphene.  \label{fig:additionalmeasurements}}
\end{figure}
Figure \ref{fig:additionalmeasurements} shows additional measurements that are included in Table 1 in the main section of the paper. These measurements are taken by the oscillator circuit, but we find that the oscillator circuit becomes often becomes unstable once the RF power is switched on, and the circuit has to be reset numerous times until it locks to the correct frequency. This leads to missing data points, however the start and end frequency still provide information of the amount of mass removed from the crystal. The problem is amended by using the vector network analyzer, alternatively one could use better holders to reject the interference from the electric field inside the plasma chamber.

\subsection*{Measurements at low power}
\begin{figure}[h!]
\centering
\includegraphics[width=\textwidth]{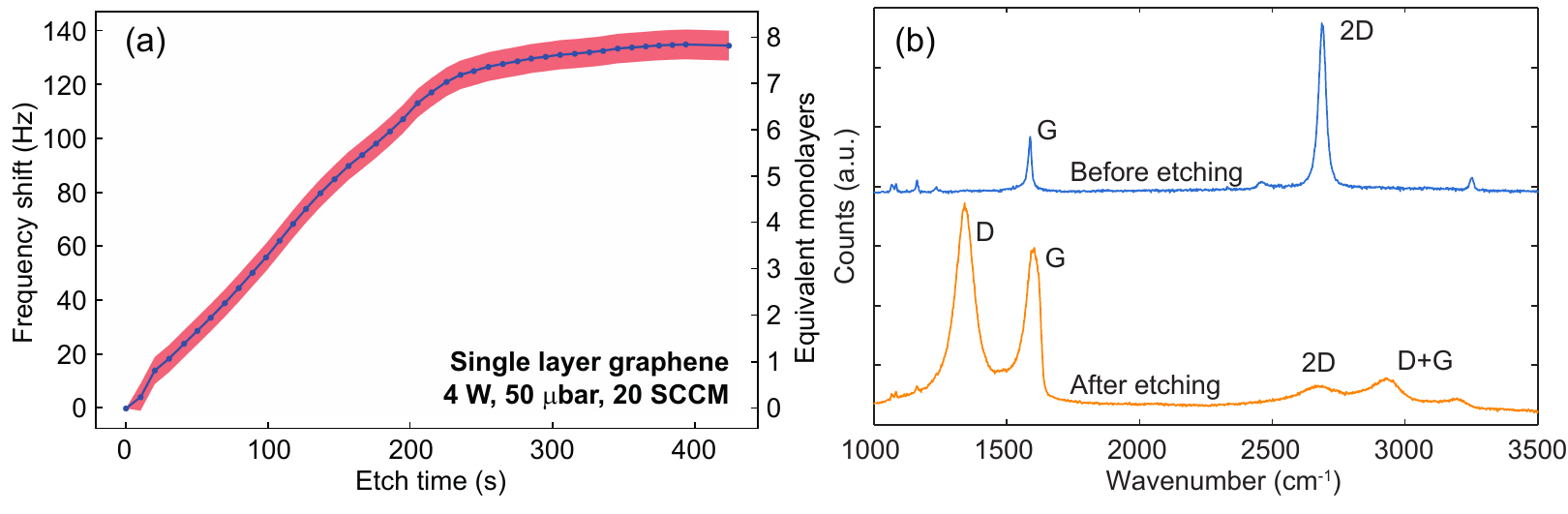}
\caption{Measurements at low etching power. (a) Etching curve of a single-layer graphene sample at low RF powers. (b) Raman spectrum before and after etching with low plasma power.  \label{fig:measlowpower}}
\end{figure}
Figure \ref{fig:measlowpower}(a) shows measurements at low RF powers of 4 W. We observe that the frequency increases and stabilizes at around 400 seconds. However, when Raman characterization was performed in Fig. \ref{fig:measlowpower}(b), is it found that the graphene is not fully removed, and the Raman spectrum bears signatures of graphene oxide \cite{C6CS00915H}. This shows that the graphene first oxidizes before it is removed from the system. In subbsequent measurements a higher power was used to ensure graphene is fully removed from the crystal. 

\subsection*{Measurements on crystal without graphene}
\begin{figure}[h!]
\centering
\includegraphics[width=\textwidth]{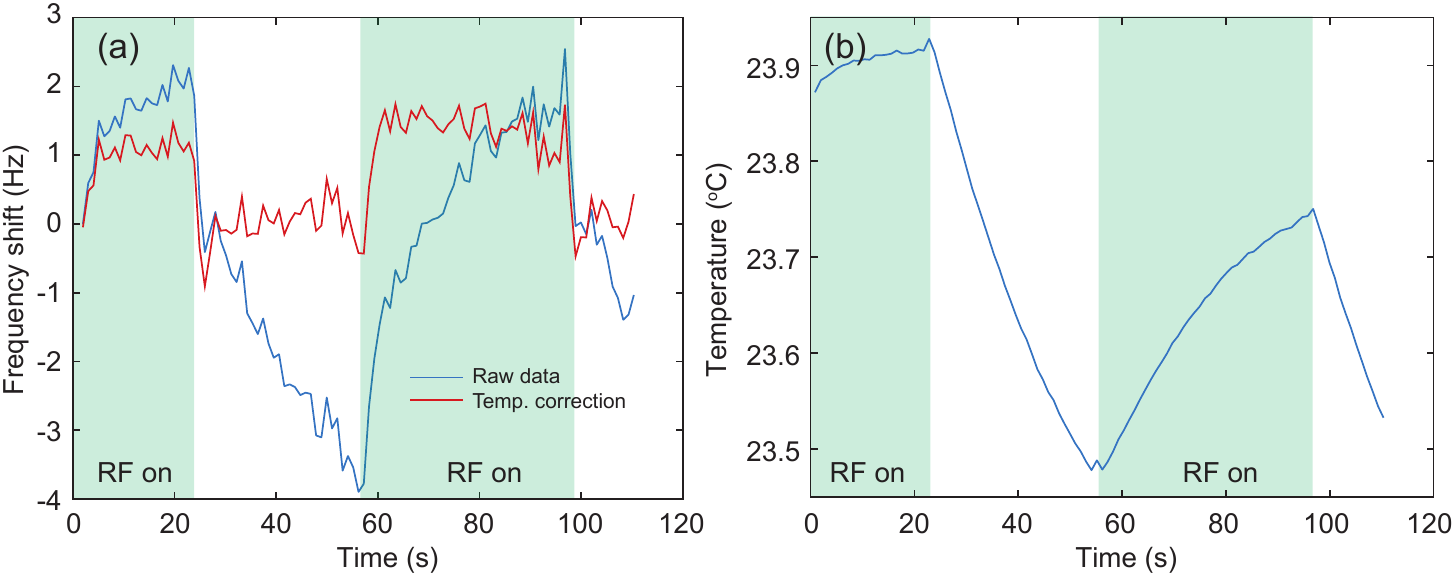}
\caption{Measurements on a crystal without graphene, using 20W, 50 $\mu$bar, 20 sccm oxygem plasma. (a) Frequency as function of time, the plasma was switched on and off. The temperature depencence was corrected for using the temperature-dependent frequency when the plasma was off. (b) Temperature as function of time. \label{fig:measclean}}
\end{figure}
Figure \ref{fig:measclean} shows measurements on a crystal without graphene that has undergone the same cleaning procedure. The RF power was switched on and off to test whether the plasma could ignite in a stable manner with this recipe and to test the stability of the frequency measurement. The frequency changes considerably during the measurement, however this can be attributed to the temperature changes in the plasma chamber. Using the frequency as function of temperature when the plasma is off, the measurement can be corrected for the temperature (red line in Fig. \ref{fig:measclean}(a)). In that case, the frequency shift is less than 0.2 Hz, which is neglectable compared to the other uncertainties in the system. 

\bibliography{dissertation}

\begin{thebibliography}{61}%
\makeatletter
\providecommand \@ifxundefined [1]{%
 \@ifx{#1\undefined}
}%
\providecommand \@ifnum [1]{%
 \ifnum #1\expandafter \@firstoftwo
 \else \expandafter \@secondoftwo
 \fi
}%
\providecommand \@ifx [1]{%
 \ifx #1\expandafter \@firstoftwo
 \else \expandafter \@secondoftwo
 \fi
}%
\providecommand \natexlab [1]{#1}%
\providecommand \enquote  [1]{``#1''}%
\providecommand \bibnamefont  [1]{#1}%
\providecommand \bibfnamefont [1]{#1}%
\providecommand \citenamefont [1]{#1}%
\providecommand \href@noop [0]{\@secondoftwo}%
\providecommand \href [0]{\begingroup \@sanitize@url \@href}%
\providecommand \@href[1]{\@@startlink{#1}\@@href}%
\providecommand \@@href[1]{\endgroup#1\@@endlink}%
\providecommand \@sanitize@url [0]{\catcode `\\12\catcode `\$12\catcode
  `\&12\catcode `\#12\catcode `\^12\catcode `\_12\catcode `\%12\relax}%
\providecommand \@@startlink[1]{}%
\providecommand \@@endlink[0]{}%
\providecommand \url  [0]{\begingroup\@sanitize@url \@url }%
\providecommand \@url [1]{\endgroup\@href {#1}{\urlprefix }}%
\providecommand \urlprefix  [0]{URL }%
\providecommand \Eprint [0]{\href }%
\providecommand \doibase [0]{http://dx.doi.org/}%
\providecommand \selectlanguage [0]{\@gobble}%
\providecommand \bibinfo  [0]{\@secondoftwo}%
\providecommand \bibfield  [0]{\@secondoftwo}%
\providecommand \translation [1]{[#1]}%
\providecommand \BibitemOpen [0]{}%
\providecommand \bibitemStop [0]{}%
\providecommand \bibitemNoStop [0]{.\EOS\space}%
\providecommand \EOS [0]{\spacefactor3000\relax}%
\providecommand \BibitemShut  [1]{\csname bibitem#1\endcsname}%
\let\auto@bib@innerbib\@empty
%</preamble>
\bibitem [{\citenamefont {Geim}\ and\ \citenamefont
  {Novoselov}(2007)}]{geim2007rise}%
  \BibitemOpen
  \bibfield  {author} {\bibinfo {author} {\bibfnamefont {Andre~K}\ \bibnamefont
  {Geim}}\ and\ \bibinfo {author} {\bibfnamefont {Konstantin~S}\ \bibnamefont
  {Novoselov}},\ }\bibfield  {title} {\enquote {\bibinfo {title} {The rise of
  graphene},}\ }\href {\doibase 10.1038/nmat1849} {\bibfield  {journal}
  {\bibinfo  {journal} {Nature Materials}\ }\textbf {\bibinfo {volume} {6}},\
  \bibinfo {pages} {183--191} (\bibinfo {year} {2007})}\BibitemShut {NoStop}%
\bibitem [{\citenamefont {Neto}\ \emph {et~al.}(2009)\citenamefont {Neto},
  \citenamefont {Guinea}, \citenamefont {Peres}, \citenamefont {Novoselov},\
  and\ \citenamefont {Geim}}]{neto2009electronic}%
  \BibitemOpen
  \bibfield  {author} {\bibinfo {author} {\bibfnamefont {AH~Castro}\
  \bibnamefont {Neto}}, \bibinfo {author} {\bibfnamefont {F}~\bibnamefont
  {Guinea}}, \bibinfo {author} {\bibfnamefont {NMR}\ \bibnamefont {Peres}},
  \bibinfo {author} {\bibfnamefont {Kostya~S}\ \bibnamefont {Novoselov}}, \
  and\ \bibinfo {author} {\bibfnamefont {Andre~K}\ \bibnamefont {Geim}},\
  }\bibfield  {title} {\enquote {\bibinfo {title} {The electronic properties of
  graphene},}\ }\href {\doibase 10.1103/RevModPhys.81.109} {\bibfield
  {journal} {\bibinfo  {journal} {Reviews of Modern Physics}\ }\textbf
  {\bibinfo {volume} {81}},\ \bibinfo {pages} {109} (\bibinfo {year}
  {2009})}\BibitemShut {NoStop}%
\bibitem [{\citenamefont {Balandin}\ \emph {et~al.}(2008)\citenamefont
  {Balandin}, \citenamefont {Ghosh}, \citenamefont {Bao}, \citenamefont
  {Calizo}, \citenamefont {Teweldebrhan}, \citenamefont {Miao},\ and\
  \citenamefont {Lau}}]{balandin2008superior}%
  \BibitemOpen
  \bibfield  {author} {\bibinfo {author} {\bibfnamefont {Alexander~A}\
  \bibnamefont {Balandin}}, \bibinfo {author} {\bibfnamefont {Suchismita}\
  \bibnamefont {Ghosh}}, \bibinfo {author} {\bibfnamefont {Wenzhong}\
  \bibnamefont {Bao}}, \bibinfo {author} {\bibfnamefont {Irene}\ \bibnamefont
  {Calizo}}, \bibinfo {author} {\bibfnamefont {Desalegne}\ \bibnamefont
  {Teweldebrhan}}, \bibinfo {author} {\bibfnamefont {Feng}\ \bibnamefont
  {Miao}}, \ and\ \bibinfo {author} {\bibfnamefont {Chun~Ning}\ \bibnamefont
  {Lau}},\ }\bibfield  {title} {\enquote {\bibinfo {title} {Superior thermal
  conductivity of single-layer graphene},}\ }\href {\doibase 10.1021/nl0731872}
  {\bibfield  {journal} {\bibinfo  {journal} {Nano Letters}\ }\textbf {\bibinfo
  {volume} {8}},\ \bibinfo {pages} {902--907} (\bibinfo {year}
  {2008})}\BibitemShut {NoStop}%
\bibitem [{\citenamefont {Pop}\ \emph {et~al.}(2012)\citenamefont {Pop},
  \citenamefont {Varshney},\ and\ \citenamefont {Roy}}]{pop2012thermal}%
  \BibitemOpen
  \bibfield  {author} {\bibinfo {author} {\bibfnamefont {Eric}\ \bibnamefont
  {Pop}}, \bibinfo {author} {\bibfnamefont {Vikas}\ \bibnamefont {Varshney}}, \
  and\ \bibinfo {author} {\bibfnamefont {Ajit~K}\ \bibnamefont {Roy}},\
  }\bibfield  {title} {\enquote {\bibinfo {title} {Thermal properties of
  graphene: Fundamentals and applications},}\ }\href {\doibase
  10.1557/mrs.2012.203} {\bibfield  {journal} {\bibinfo  {journal} {MRS
  Bulletin}\ }\textbf {\bibinfo {volume} {37}},\ \bibinfo {pages} {1273--1281}
  (\bibinfo {year} {2012})}\BibitemShut {NoStop}%
\bibitem [{\citenamefont {Dolleman}\ \emph
  {et~al.}(2017{\natexlab{a}})\citenamefont {Dolleman}, \citenamefont {Houri},
  \citenamefont {Davidovikj}, \citenamefont {Cartamil-Bueno}, \citenamefont
  {Blanter}, \citenamefont {van~der Zant},\ and\ \citenamefont
  {Steeneken}}]{dolleman2017optomechanics}%
  \BibitemOpen
  \bibfield  {author} {\bibinfo {author} {\bibfnamefont {Robin~J}\ \bibnamefont
  {Dolleman}}, \bibinfo {author} {\bibfnamefont {Samer}\ \bibnamefont {Houri}},
  \bibinfo {author} {\bibfnamefont {Dejan}\ \bibnamefont {Davidovikj}},
  \bibinfo {author} {\bibfnamefont {Santiago~J}\ \bibnamefont
  {Cartamil-Bueno}}, \bibinfo {author} {\bibfnamefont {Yaroslav~M}\
  \bibnamefont {Blanter}}, \bibinfo {author} {\bibfnamefont {Herre~SJ}\
  \bibnamefont {van~der Zant}}, \ and\ \bibinfo {author} {\bibfnamefont
  {Peter~G}\ \bibnamefont {Steeneken}},\ }\bibfield  {title} {\enquote
  {\bibinfo {title} {Optomechanics for thermal characterization of suspended
  graphene},}\ }\href {\doibase 10.1103/PhysRevB.96.165421} {\bibfield
  {journal} {\bibinfo  {journal} {Physical Review B}\ }\textbf {\bibinfo
  {volume} {96}},\ \bibinfo {pages} {165421} (\bibinfo {year}
  {2017}{\natexlab{a}})}\BibitemShut {NoStop}%
\bibitem [{\citenamefont {Lee}\ \emph {et~al.}(2008)\citenamefont {Lee},
  \citenamefont {Wei}, \citenamefont {Kysar},\ and\ \citenamefont
  {Hone}}]{lee2008measurement}%
  \BibitemOpen
  \bibfield  {author} {\bibinfo {author} {\bibfnamefont {Changgu}\ \bibnamefont
  {Lee}}, \bibinfo {author} {\bibfnamefont {Xiaoding}\ \bibnamefont {Wei}},
  \bibinfo {author} {\bibfnamefont {Jeffrey~W.}\ \bibnamefont {Kysar}}, \ and\
  \bibinfo {author} {\bibfnamefont {James}\ \bibnamefont {Hone}},\ }\bibfield
  {title} {\enquote {\bibinfo {title} {Measurement of the elastic properties
  and intrinsic strength of monolayer graphene},}\ }\href {\doibase
  10.1126/science.1157996} {\bibfield  {journal} {\bibinfo  {journal}
  {Science}\ }\textbf {\bibinfo {volume} {321}},\ \bibinfo {pages} {385--388}
  (\bibinfo {year} {2008})}\BibitemShut {NoStop}%
\bibitem [{\citenamefont {Davidovikj}\ \emph
  {et~al.}(2017{\natexlab{a}})\citenamefont {Davidovikj}, \citenamefont
  {Alijani}, \citenamefont {Cartamil-Bueno}, \citenamefont {van~der Zant},
  \citenamefont {Amabili},\ and\ \citenamefont
  {Steeneken}}]{davidovikj2017nonlinear}%
  \BibitemOpen
  \bibfield  {author} {\bibinfo {author} {\bibfnamefont {D.}~\bibnamefont
  {Davidovikj}}, \bibinfo {author} {\bibfnamefont {F.}~\bibnamefont {Alijani}},
  \bibinfo {author} {\bibfnamefont {S.~J.}\ \bibnamefont {Cartamil-Bueno}},
  \bibinfo {author} {\bibfnamefont {H.~S.~J.}\ \bibnamefont {van~der Zant}},
  \bibinfo {author} {\bibfnamefont {M.}~\bibnamefont {Amabili}}, \ and\
  \bibinfo {author} {\bibfnamefont {P.~G.}\ \bibnamefont {Steeneken}},\
  }\bibfield  {title} {\enquote {\bibinfo {title} {Nonlinear dynamic
  characterization of two-dimensional materials},}\ }\href {\doibase
  10.1038/s41467-017-01351-4} {\bibfield  {journal} {\bibinfo  {journal}
  {Nature Communications}\ }\textbf {\bibinfo {volume} {8}},\ \bibinfo {pages}
  {1253} (\bibinfo {year} {2017}{\natexlab{a}})}\BibitemShut {NoStop}%
\bibitem [{\citenamefont {Sajadi}\ \emph {et~al.}(2017)\citenamefont {Sajadi},
  \citenamefont {Alijani}, \citenamefont {Davidovikj}, \citenamefont {Goosen},
  \citenamefont {Steeneken},\ and\ \citenamefont {van
  Keulen}}]{sajadi2017experimental}%
  \BibitemOpen
  \bibfield  {author} {\bibinfo {author} {\bibfnamefont {Banafsheh}\
  \bibnamefont {Sajadi}}, \bibinfo {author} {\bibfnamefont {Farbod}\
  \bibnamefont {Alijani}}, \bibinfo {author} {\bibfnamefont {Dejan}\
  \bibnamefont {Davidovikj}}, \bibinfo {author} {\bibfnamefont {Johannes}\
  \bibnamefont {Goosen}}, \bibinfo {author} {\bibfnamefont {Peter~G}\
  \bibnamefont {Steeneken}}, \ and\ \bibinfo {author} {\bibfnamefont {Fred}\
  \bibnamefont {van Keulen}},\ }\bibfield  {title} {\enquote {\bibinfo {title}
  {Experimental characterization of graphene by electrostatic resonance
  frequency tuning},}\ }\href {\doibase 10.1063/1.4999682} {\bibfield
  {journal} {\bibinfo  {journal} {Journal of Applied Physics}\ }\textbf
  {\bibinfo {volume} {122}},\ \bibinfo {pages} {234302} (\bibinfo {year}
  {2017})}\BibitemShut {NoStop}%
\bibitem [{\citenamefont {Song}\ \emph {et~al.}(2011)\citenamefont {Song},
  \citenamefont {Oksanen}, \citenamefont {Sillanpaa}, \citenamefont
  {Craighead}, \citenamefont {Parpia},\ and\ \citenamefont
  {Hakonen}}]{song2011stamp}%
  \BibitemOpen
  \bibfield  {author} {\bibinfo {author} {\bibfnamefont {Xuefeng}\ \bibnamefont
  {Song}}, \bibinfo {author} {\bibfnamefont {Mika}\ \bibnamefont {Oksanen}},
  \bibinfo {author} {\bibfnamefont {Mika~A}\ \bibnamefont {Sillanpaa}},
  \bibinfo {author} {\bibfnamefont {HG}~\bibnamefont {Craighead}}, \bibinfo
  {author} {\bibfnamefont {JM}~\bibnamefont {Parpia}}, \ and\ \bibinfo {author}
  {\bibfnamefont {Pertti~J}\ \bibnamefont {Hakonen}},\ }\bibfield  {title}
  {\enquote {\bibinfo {title} {Stamp transferred suspended graphene mechanical
  resonators for radio frequency electrical readout},}\ }\href {\doibase
  10.1021/nl203305q} {\bibfield  {journal} {\bibinfo  {journal} {Nano Letters}\
  }\textbf {\bibinfo {volume} {12}},\ \bibinfo {pages} {198--202} (\bibinfo
  {year} {2011})}\BibitemShut {NoStop}%
\bibitem [{\citenamefont {Chen}\ \emph
  {et~al.}(2013{\natexlab{a}})\citenamefont {Chen}, \citenamefont {Lee},
  \citenamefont {Deshpande}, \citenamefont {Lee}, \citenamefont {Lekas},
  \citenamefont {Shepard},\ and\ \citenamefont {Hone}}]{chen2013graphene}%
  \BibitemOpen
  \bibfield  {author} {\bibinfo {author} {\bibfnamefont {Changyao}\
  \bibnamefont {Chen}}, \bibinfo {author} {\bibfnamefont {Sunwoo}\ \bibnamefont
  {Lee}}, \bibinfo {author} {\bibfnamefont {Vikram~V}\ \bibnamefont
  {Deshpande}}, \bibinfo {author} {\bibfnamefont {Gwan-Hyoung}\ \bibnamefont
  {Lee}}, \bibinfo {author} {\bibfnamefont {Michael}\ \bibnamefont {Lekas}},
  \bibinfo {author} {\bibfnamefont {Kenneth}\ \bibnamefont {Shepard}}, \ and\
  \bibinfo {author} {\bibfnamefont {James}\ \bibnamefont {Hone}},\ }\bibfield
  {title} {\enquote {\bibinfo {title} {Graphene mechanical oscillators with
  tunable frequency},}\ }\href {\doibase 10.1038/nnano.2013.232} {\bibfield
  {journal} {\bibinfo  {journal} {Nature Nanotechnology}\ }\textbf {\bibinfo
  {volume} {8}},\ \bibinfo {pages} {923--927} (\bibinfo {year}
  {2013}{\natexlab{a}})}\BibitemShut {NoStop}%
\bibitem [{\citenamefont {Lemme}\ \emph {et~al.}(2007)\citenamefont {Lemme},
  \citenamefont {Echtermeyer}, \citenamefont {Baus},\ and\ \citenamefont
  {Kurz}}]{lemme2007graphene}%
  \BibitemOpen
  \bibfield  {author} {\bibinfo {author} {\bibfnamefont {Max~C}\ \bibnamefont
  {Lemme}}, \bibinfo {author} {\bibfnamefont {Tim~J}\ \bibnamefont
  {Echtermeyer}}, \bibinfo {author} {\bibfnamefont {Matthias}\ \bibnamefont
  {Baus}}, \ and\ \bibinfo {author} {\bibfnamefont {Heinrich}\ \bibnamefont
  {Kurz}},\ }\bibfield  {title} {\enquote {\bibinfo {title} {A graphene
  field-effect device},}\ }\href@noop {} {\bibfield  {journal} {\bibinfo
  {journal} {IEEE Electron Device Letters}\ }\textbf {\bibinfo {volume} {28}},\
  \bibinfo {pages} {282--284} (\bibinfo {year} {2007})}\BibitemShut {NoStop}%
\bibitem [{\citenamefont {Cartamil-Bueno}\ \emph {et~al.}(2018)\citenamefont
  {Cartamil-Bueno}, \citenamefont {Davidovikj}, \citenamefont {Centeno},
  \citenamefont {Zurutuza}, \citenamefont {van~der Zant}, \citenamefont
  {Steeneken},\ and\ \citenamefont {Houri}}]{gimod}%
  \BibitemOpen
  \bibfield  {author} {\bibinfo {author} {\bibfnamefont {Santiago~J.}\
  \bibnamefont {Cartamil-Bueno}}, \bibinfo {author} {\bibfnamefont {Dejan}\
  \bibnamefont {Davidovikj}}, \bibinfo {author} {\bibfnamefont {Alba}\
  \bibnamefont {Centeno}}, \bibinfo {author} {\bibfnamefont {Amaia}\
  \bibnamefont {Zurutuza}}, \bibinfo {author} {\bibfnamefont {Herre S.~J.}\
  \bibnamefont {van~der Zant}}, \bibinfo {author} {\bibfnamefont {Peter~G.}\
  \bibnamefont {Steeneken}}, \ and\ \bibinfo {author} {\bibfnamefont {Samer}\
  \bibnamefont {Houri}},\ }\bibfield  {title} {\enquote {\bibinfo {title}
  {Graphene mechanical pixels for interferometric modulator displays},}\ }\href
  {\doibase 10.1038/s41467-018-07230-w} {\bibfield  {journal} {\bibinfo
  {journal} {Nature Communications}\ }\textbf {\bibinfo {volume} {9}},\
  \bibinfo {pages} {4837} (\bibinfo {year} {2018})}\BibitemShut {NoStop}%
\bibitem [{\citenamefont {Dauber}\ \emph {et~al.}(2015)\citenamefont {Dauber},
  \citenamefont {Sagade}, \citenamefont {Oellers}, \citenamefont {Watanabe},
  \citenamefont {Taniguchi}, \citenamefont {Neumaier},\ and\ \citenamefont
  {Stampfer}}]{dauber2015ultra}%
  \BibitemOpen
  \bibfield  {author} {\bibinfo {author} {\bibfnamefont {Jan}\ \bibnamefont
  {Dauber}}, \bibinfo {author} {\bibfnamefont {Abhay~A}\ \bibnamefont
  {Sagade}}, \bibinfo {author} {\bibfnamefont {Martin}\ \bibnamefont
  {Oellers}}, \bibinfo {author} {\bibfnamefont {Kenji}\ \bibnamefont
  {Watanabe}}, \bibinfo {author} {\bibfnamefont {Takashi}\ \bibnamefont
  {Taniguchi}}, \bibinfo {author} {\bibfnamefont {Daniel}\ \bibnamefont
  {Neumaier}}, \ and\ \bibinfo {author} {\bibfnamefont {Christoph}\
  \bibnamefont {Stampfer}},\ }\bibfield  {title} {\enquote {\bibinfo {title}
  {Ultra-sensitive {Hall} sensors based on graphene encapsulated in hexagonal
  boron nitride},}\ }\href@noop {} {\bibfield  {journal} {\bibinfo  {journal}
  {Applied Physics Letters}\ }\textbf {\bibinfo {volume} {106}},\ \bibinfo
  {pages} {193501} (\bibinfo {year} {2015})}\BibitemShut {NoStop}%
\bibitem [{\citenamefont {Smith}\ \emph {et~al.}(2017)\citenamefont {Smith},
  \citenamefont {Elgammal}, \citenamefont {Fan}, \citenamefont {Lemme},
  \citenamefont {Delin}, \citenamefont {R{\aa}sander}, \citenamefont
  {Bergqvist}, \citenamefont {Schr{\"o}der}, \citenamefont {Fischer},
  \citenamefont {Niklaus} \emph {et~al.}}]{smith2017graphene}%
  \BibitemOpen
  \bibfield  {author} {\bibinfo {author} {\bibfnamefont {Anderson~D}\
  \bibnamefont {Smith}}, \bibinfo {author} {\bibfnamefont {Karim}\ \bibnamefont
  {Elgammal}}, \bibinfo {author} {\bibfnamefont {Xuge}\ \bibnamefont {Fan}},
  \bibinfo {author} {\bibfnamefont {Max~C}\ \bibnamefont {Lemme}}, \bibinfo
  {author} {\bibfnamefont {Anna}\ \bibnamefont {Delin}}, \bibinfo {author}
  {\bibfnamefont {Mikael}\ \bibnamefont {R{\aa}sander}}, \bibinfo {author}
  {\bibfnamefont {Lars}\ \bibnamefont {Bergqvist}}, \bibinfo {author}
  {\bibfnamefont {Stephan}\ \bibnamefont {Schr{\"o}der}}, \bibinfo {author}
  {\bibfnamefont {Andreas~C}\ \bibnamefont {Fischer}}, \bibinfo {author}
  {\bibfnamefont {Frank}\ \bibnamefont {Niklaus}},  \emph {et~al.},\ }\bibfield
   {title} {\enquote {\bibinfo {title} {Graphene-based co 2 sensing and its
  cross-sensitivity with humidity},}\ }\href@noop {} {\bibfield  {journal}
  {\bibinfo  {journal} {RSC Advances}\ }\textbf {\bibinfo {volume} {7}},\
  \bibinfo {pages} {22329--22339} (\bibinfo {year} {2017})}\BibitemShut
  {NoStop}%
\bibitem [{\citenamefont {Ricciardella}\ \emph {et~al.}(2017)\citenamefont
  {Ricciardella}, \citenamefont {Vollebregt}, \citenamefont {Polichetti},
  \citenamefont {Miscuglio}, \citenamefont {Alfano}, \citenamefont {Miglietta},
  \citenamefont {Massera}, \citenamefont {Di~Francia},\ and\ \citenamefont
  {Sarro}}]{ricciardella2017effects}%
  \BibitemOpen
  \bibfield  {author} {\bibinfo {author} {\bibfnamefont {Filiberto}\
  \bibnamefont {Ricciardella}}, \bibinfo {author} {\bibfnamefont {Sten}\
  \bibnamefont {Vollebregt}}, \bibinfo {author} {\bibfnamefont {Tiziana}\
  \bibnamefont {Polichetti}}, \bibinfo {author} {\bibfnamefont {Mario}\
  \bibnamefont {Miscuglio}}, \bibinfo {author} {\bibfnamefont {Brigida}\
  \bibnamefont {Alfano}}, \bibinfo {author} {\bibfnamefont {Maria~L}\
  \bibnamefont {Miglietta}}, \bibinfo {author} {\bibfnamefont {Ettore}\
  \bibnamefont {Massera}}, \bibinfo {author} {\bibfnamefont {Girolamo}\
  \bibnamefont {Di~Francia}}, \ and\ \bibinfo {author} {\bibfnamefont
  {Pasqualina~M}\ \bibnamefont {Sarro}},\ }\bibfield  {title} {\enquote
  {\bibinfo {title} {Effects of graphene defects on gas sensing properties
  towards no 2 detection},}\ }\href@noop {} {\bibfield  {journal} {\bibinfo
  {journal} {Nanoscale}\ }\textbf {\bibinfo {volume} {9}},\ \bibinfo {pages}
  {6085--6093} (\bibinfo {year} {2017})}\BibitemShut {NoStop}%
\bibitem [{\citenamefont {Smith}\ \emph {et~al.}(2013)\citenamefont {Smith},
  \citenamefont {Vaziri}, \citenamefont {Niklaus}, \citenamefont {Fischer},
  \citenamefont {Sterner}, \citenamefont {Delin}, \citenamefont {{\"O}stling},\
  and\ \citenamefont {Lemme}}]{smith2013pressure}%
  \BibitemOpen
  \bibfield  {author} {\bibinfo {author} {\bibfnamefont {Anderson~D}\
  \bibnamefont {Smith}}, \bibinfo {author} {\bibfnamefont {Sam}\ \bibnamefont
  {Vaziri}}, \bibinfo {author} {\bibfnamefont {Frank}\ \bibnamefont {Niklaus}},
  \bibinfo {author} {\bibfnamefont {Andreas~C}\ \bibnamefont {Fischer}},
  \bibinfo {author} {\bibfnamefont {Mikael}\ \bibnamefont {Sterner}}, \bibinfo
  {author} {\bibfnamefont {Anna}\ \bibnamefont {Delin}}, \bibinfo {author}
  {\bibfnamefont {Mikael}\ \bibnamefont {{\"O}stling}}, \ and\ \bibinfo
  {author} {\bibfnamefont {Max~C}\ \bibnamefont {Lemme}},\ }\bibfield  {title}
  {\enquote {\bibinfo {title} {Pressure sensors based on suspended graphene
  membranes},}\ }\href {\doibase 10.1016/j.sse.2013.04.019} {\bibfield
  {journal} {\bibinfo  {journal} {Solid-State Electronics}\ }\textbf {\bibinfo
  {volume} {88}},\ \bibinfo {pages} {89--94} (\bibinfo {year}
  {2013})}\BibitemShut {NoStop}%
\bibitem [{\citenamefont {Dolleman}\ \emph {et~al.}(2016)\citenamefont
  {Dolleman}, \citenamefont {Davidovikj}, \citenamefont {Cartamil-Bueno},
  \citenamefont {van~der Zant},\ and\ \citenamefont
  {Steeneken}}]{dolleman2015graphene}%
  \BibitemOpen
  \bibfield  {author} {\bibinfo {author} {\bibfnamefont {Robin~J}\ \bibnamefont
  {Dolleman}}, \bibinfo {author} {\bibfnamefont {Dejan}\ \bibnamefont
  {Davidovikj}}, \bibinfo {author} {\bibfnamefont {Santiago~J}\ \bibnamefont
  {Cartamil-Bueno}}, \bibinfo {author} {\bibfnamefont {Herre~SJ}\ \bibnamefont
  {van~der Zant}}, \ and\ \bibinfo {author} {\bibfnamefont {Peter~G}\
  \bibnamefont {Steeneken}},\ }\bibfield  {title} {\enquote {\bibinfo {title}
  {Graphene squeeze-film pressure sensors},}\ }\href {\doibase
  10.1021/acs.nanolett.5b04251} {\bibfield  {journal} {\bibinfo  {journal}
  {Nano Letters}\ }\textbf {\bibinfo {volume} {16}},\ \bibinfo {pages}
  {568--571} (\bibinfo {year} {2016})}\BibitemShut {NoStop}%
\bibitem [{\citenamefont {Dolleman}\ \emph
  {et~al.}(2017{\natexlab{b}})\citenamefont {Dolleman}, \citenamefont
  {Cartamil-Bueno}, \citenamefont {van~der Zant},\ and\ \citenamefont
  {Steeneken}}]{dolleman2016graphene}%
  \BibitemOpen
  \bibfield  {author} {\bibinfo {author} {\bibfnamefont {Robin~J}\ \bibnamefont
  {Dolleman}}, \bibinfo {author} {\bibfnamefont {Santiago~J}\ \bibnamefont
  {Cartamil-Bueno}}, \bibinfo {author} {\bibfnamefont {Herre S~J}\ \bibnamefont
  {van~der Zant}}, \ and\ \bibinfo {author} {\bibfnamefont {Peter~G}\
  \bibnamefont {Steeneken}},\ }\bibfield  {title} {\enquote {\bibinfo {title}
  {Graphene gas osmometers},}\ }\href {\doibase 10.1088/2053-1583/4/1/011002}
  {\bibfield  {journal} {\bibinfo  {journal} {2D Materials}\ }\textbf {\bibinfo
  {volume} {4}},\ \bibinfo {pages} {011002} (\bibinfo {year}
  {2017}{\natexlab{b}})}\BibitemShut {NoStop}%
\bibitem [{\citenamefont {Vollebregt}\ \emph {et~al.}(2017)\citenamefont
  {Vollebregt}, \citenamefont {Dolleman}, \citenamefont {van~der Zant},
  \citenamefont {Steeneken},\ and\ \citenamefont
  {Sarro}}]{vollebregt2017suspended}%
  \BibitemOpen
  \bibfield  {author} {\bibinfo {author} {\bibfnamefont {Sten}\ \bibnamefont
  {Vollebregt}}, \bibinfo {author} {\bibfnamefont {Robin~J}\ \bibnamefont
  {Dolleman}}, \bibinfo {author} {\bibfnamefont {Herre S~J}\ \bibnamefont
  {van~der Zant}}, \bibinfo {author} {\bibfnamefont {Peter~G}\ \bibnamefont
  {Steeneken}}, \ and\ \bibinfo {author} {\bibfnamefont {P~M}\ \bibnamefont
  {Sarro}},\ }\bibfield  {title} {\enquote {\bibinfo {title} {Suspended
  graphene beams with tunable gap for squeeze-film pressure sensing},}\ }in\
  \href {\doibase 10.1109/TRANSDUCERS.2017.7994162} {\emph {\bibinfo
  {booktitle} {Transducers 2017, The 19th International Conference on
  Solid-State Sensors, Actuators and Microsystems}}}\ (\bibinfo {organization}
  {IEEE},\ \bibinfo {year} {2017})\ pp.\ \bibinfo {pages}
  {770--773}\BibitemShut {NoStop}%
\bibitem [{\citenamefont {Davidovikj}\ \emph
  {et~al.}(2017{\natexlab{b}})\citenamefont {Davidovikj}, \citenamefont
  {Scheepers}, \citenamefont {van~der Zant},\ and\ \citenamefont
  {Steeneken}}]{davidovikj2017static}%
  \BibitemOpen
  \bibfield  {author} {\bibinfo {author} {\bibfnamefont {Dejan}\ \bibnamefont
  {Davidovikj}}, \bibinfo {author} {\bibfnamefont {Paul~H}\ \bibnamefont
  {Scheepers}}, \bibinfo {author} {\bibfnamefont {Herre S~J}\ \bibnamefont
  {van~der Zant}}, \ and\ \bibinfo {author} {\bibfnamefont {Peter~G}\
  \bibnamefont {Steeneken}},\ }\bibfield  {title} {\enquote {\bibinfo {title}
  {Static capacitive pressure sensing using a single graphene drum},}\ }\href
  {\doibase 10.1021/acsami.7b17487} {\bibfield  {journal} {\bibinfo  {journal}
  {ACS Applied Materials \& Interfaces}\ }\textbf {\bibinfo {volume} {9}},\
  \bibinfo {pages} {43205--43210} (\bibinfo {year}
  {2017}{\natexlab{b}})}\BibitemShut {NoStop}%
\bibitem [{\citenamefont {Suk}\ \emph {et~al.}(2011)\citenamefont {Suk},
  \citenamefont {Kitt}, \citenamefont {Magnuson}, \citenamefont {Hao},
  \citenamefont {Ahmed}, \citenamefont {An}, \citenamefont {Swan},
  \citenamefont {Goldberg},\ and\ \citenamefont {Ruoff}}]{suk2011transfer}%
  \BibitemOpen
  \bibfield  {author} {\bibinfo {author} {\bibfnamefont {Ji~Won}\ \bibnamefont
  {Suk}}, \bibinfo {author} {\bibfnamefont {Alexander}\ \bibnamefont {Kitt}},
  \bibinfo {author} {\bibfnamefont {Carl~W}\ \bibnamefont {Magnuson}}, \bibinfo
  {author} {\bibfnamefont {Yufeng}\ \bibnamefont {Hao}}, \bibinfo {author}
  {\bibfnamefont {Samir}\ \bibnamefont {Ahmed}}, \bibinfo {author}
  {\bibfnamefont {Jinho}\ \bibnamefont {An}}, \bibinfo {author} {\bibfnamefont
  {Anna~K}\ \bibnamefont {Swan}}, \bibinfo {author} {\bibfnamefont {Bennett~B}\
  \bibnamefont {Goldberg}}, \ and\ \bibinfo {author} {\bibfnamefont {Rodney~S}\
  \bibnamefont {Ruoff}},\ }\bibfield  {title} {\enquote {\bibinfo {title}
  {Transfer of {CVD}-grown monolayer graphene onto arbitrary substrates},}\
  }\href {\doibase 10.1021/nn201207c} {\bibfield  {journal} {\bibinfo
  {journal} {ACS Nano}\ }\textbf {\bibinfo {volume} {5}},\ \bibinfo {pages}
  {6916--6924} (\bibinfo {year} {2011})}\BibitemShut {NoStop}%
\bibitem [{\citenamefont {Li}\ \emph {et~al.}(2009{\natexlab{a}})\citenamefont
  {Li}, \citenamefont {Cai}, \citenamefont {An}, \citenamefont {Kim},
  \citenamefont {Nah}, \citenamefont {Yang}, \citenamefont {Piner},
  \citenamefont {Velamakanni}, \citenamefont {Jung}, \citenamefont {Tutuc}
  \emph {et~al.}}]{li2009large}%
  \BibitemOpen
  \bibfield  {author} {\bibinfo {author} {\bibfnamefont {Xuesong}\ \bibnamefont
  {Li}}, \bibinfo {author} {\bibfnamefont {Weiwei}\ \bibnamefont {Cai}},
  \bibinfo {author} {\bibfnamefont {Jinho}\ \bibnamefont {An}}, \bibinfo
  {author} {\bibfnamefont {Seyoung}\ \bibnamefont {Kim}}, \bibinfo {author}
  {\bibfnamefont {Junghyo}\ \bibnamefont {Nah}}, \bibinfo {author}
  {\bibfnamefont {Dongxing}\ \bibnamefont {Yang}}, \bibinfo {author}
  {\bibfnamefont {Richard}\ \bibnamefont {Piner}}, \bibinfo {author}
  {\bibfnamefont {Aruna}\ \bibnamefont {Velamakanni}}, \bibinfo {author}
  {\bibfnamefont {Inhwa}\ \bibnamefont {Jung}}, \bibinfo {author}
  {\bibfnamefont {Emanuel}\ \bibnamefont {Tutuc}},  \emph {et~al.},\ }\bibfield
   {title} {\enquote {\bibinfo {title} {Large-area synthesis of high-quality
  and uniform graphene films on copper foils},}\ }\href@noop {} {\bibfield
  {journal} {\bibinfo  {journal} {Science}\ }\textbf {\bibinfo {volume}
  {324}},\ \bibinfo {pages} {1312--1314} (\bibinfo {year}
  {2009}{\natexlab{a}})}\BibitemShut {NoStop}%
\bibitem [{\citenamefont {Li}\ \emph {et~al.}(2009{\natexlab{b}})\citenamefont
  {Li}, \citenamefont {Zhu}, \citenamefont {Cai}, \citenamefont {Borysiak},
  \citenamefont {Han}, \citenamefont {Chen}, \citenamefont {Piner},
  \citenamefont {Colombo},\ and\ \citenamefont {Ruoff}}]{li2009transfer}%
  \BibitemOpen
  \bibfield  {author} {\bibinfo {author} {\bibfnamefont {Xuesong}\ \bibnamefont
  {Li}}, \bibinfo {author} {\bibfnamefont {Yanwu}\ \bibnamefont {Zhu}},
  \bibinfo {author} {\bibfnamefont {Weiwei}\ \bibnamefont {Cai}}, \bibinfo
  {author} {\bibfnamefont {Mark}\ \bibnamefont {Borysiak}}, \bibinfo {author}
  {\bibfnamefont {Boyang}\ \bibnamefont {Han}}, \bibinfo {author}
  {\bibfnamefont {David}\ \bibnamefont {Chen}}, \bibinfo {author}
  {\bibfnamefont {Richard~D}\ \bibnamefont {Piner}}, \bibinfo {author}
  {\bibfnamefont {Luigi}\ \bibnamefont {Colombo}}, \ and\ \bibinfo {author}
  {\bibfnamefont {Rodney~S}\ \bibnamefont {Ruoff}},\ }\bibfield  {title}
  {\enquote {\bibinfo {title} {Transfer of large-area graphene films for
  high-performance transparent conductive electrodes},}\ }\href@noop {}
  {\bibfield  {journal} {\bibinfo  {journal} {Nano letters}\ }\textbf {\bibinfo
  {volume} {9}},\ \bibinfo {pages} {4359--4363} (\bibinfo {year}
  {2009}{\natexlab{b}})}\BibitemShut {NoStop}%
\bibitem [{\citenamefont {Lee}\ \emph {et~al.}(2010)\citenamefont {Lee},
  \citenamefont {Bae}, \citenamefont {Jang}, \citenamefont {Jang},
  \citenamefont {Zhu}, \citenamefont {Sim}, \citenamefont {Song}, \citenamefont
  {Hong},\ and\ \citenamefont {Ahn}}]{lee2010wafer}%
  \BibitemOpen
  \bibfield  {author} {\bibinfo {author} {\bibfnamefont {Youngbin}\
  \bibnamefont {Lee}}, \bibinfo {author} {\bibfnamefont {Sukang}\ \bibnamefont
  {Bae}}, \bibinfo {author} {\bibfnamefont {Houk}\ \bibnamefont {Jang}},
  \bibinfo {author} {\bibfnamefont {Sukjae}\ \bibnamefont {Jang}}, \bibinfo
  {author} {\bibfnamefont {Shou-En}\ \bibnamefont {Zhu}}, \bibinfo {author}
  {\bibfnamefont {Sung~Hyun}\ \bibnamefont {Sim}}, \bibinfo {author}
  {\bibfnamefont {Young~Il}\ \bibnamefont {Song}}, \bibinfo {author}
  {\bibfnamefont {Byung~Hee}\ \bibnamefont {Hong}}, \ and\ \bibinfo {author}
  {\bibfnamefont {Jong-Hyun}\ \bibnamefont {Ahn}},\ }\bibfield  {title}
  {\enquote {\bibinfo {title} {Wafer-scale synthesis and transfer of graphene
  films},}\ }\href@noop {} {\bibfield  {journal} {\bibinfo  {journal} {Nano
  Lletters}\ }\textbf {\bibinfo {volume} {10}},\ \bibinfo {pages} {490--493}
  (\bibinfo {year} {2010})}\BibitemShut {NoStop}%
\bibitem [{\citenamefont {Chen}\ \emph
  {et~al.}(2013{\natexlab{b}})\citenamefont {Chen}, \citenamefont {Liu},
  \citenamefont {Zheng}, \citenamefont {Xing}, \citenamefont {Yan},
  \citenamefont {Chen},\ and\ \citenamefont {Tian}}]{chen2013high}%
  \BibitemOpen
  \bibfield  {author} {\bibinfo {author} {\bibfnamefont {Xu-Dong}\ \bibnamefont
  {Chen}}, \bibinfo {author} {\bibfnamefont {Zhi-Bo}\ \bibnamefont {Liu}},
  \bibinfo {author} {\bibfnamefont {Chao-Yi}\ \bibnamefont {Zheng}}, \bibinfo
  {author} {\bibfnamefont {Fei}\ \bibnamefont {Xing}}, \bibinfo {author}
  {\bibfnamefont {Xiao-Qing}\ \bibnamefont {Yan}}, \bibinfo {author}
  {\bibfnamefont {Yongsheng}\ \bibnamefont {Chen}}, \ and\ \bibinfo {author}
  {\bibfnamefont {Jian-Guo}\ \bibnamefont {Tian}},\ }\bibfield  {title}
  {\enquote {\bibinfo {title} {High-quality and efficient transfer of
  large-area graphene films onto different substrates},}\ }\href@noop {}
  {\bibfield  {journal} {\bibinfo  {journal} {Carbon}\ }\textbf {\bibinfo
  {volume} {56}},\ \bibinfo {pages} {271--278} (\bibinfo {year}
  {2013}{\natexlab{b}})}\BibitemShut {NoStop}%
\bibitem [{\citenamefont {Zande}\ \emph {et~al.}(2010)\citenamefont {Zande},
  \citenamefont {Barton}, \citenamefont {Alden}, \citenamefont {Ruiz-Vargas},
  \citenamefont {Whitney}, \citenamefont {Pham}, \citenamefont {Park},
  \citenamefont {Parpia}, \citenamefont {Craighead},\ and\ \citenamefont
  {McEuen}}]{zande2010large}%
  \BibitemOpen
  \bibfield  {author} {\bibinfo {author} {\bibfnamefont {Arend M van~der}\
  \bibnamefont {Zande}}, \bibinfo {author} {\bibfnamefont {Robert~A}\
  \bibnamefont {Barton}}, \bibinfo {author} {\bibfnamefont {Jonathan~S}\
  \bibnamefont {Alden}}, \bibinfo {author} {\bibfnamefont {Carlos~S}\
  \bibnamefont {Ruiz-Vargas}}, \bibinfo {author} {\bibfnamefont {William~S}\
  \bibnamefont {Whitney}}, \bibinfo {author} {\bibfnamefont {Phi~HQ}\
  \bibnamefont {Pham}}, \bibinfo {author} {\bibfnamefont {Jiwoong}\
  \bibnamefont {Park}}, \bibinfo {author} {\bibfnamefont {Jeevak~M}\
  \bibnamefont {Parpia}}, \bibinfo {author} {\bibfnamefont {Harold~G}\
  \bibnamefont {Craighead}}, \ and\ \bibinfo {author} {\bibfnamefont {Paul~L}\
  \bibnamefont {McEuen}},\ }\bibfield  {title} {\enquote {\bibinfo {title}
  {Large-scale arrays of single-layer graphene resonators},}\ }\href {\doibase
  10.1021/nl102713c} {\bibfield  {journal} {\bibinfo  {journal} {Nano Letters}\
  }\textbf {\bibinfo {volume} {10}},\ \bibinfo {pages} {4869--4873} (\bibinfo
  {year} {2010})}\BibitemShut {NoStop}%
\bibitem [{\citenamefont {Her}\ \emph {et~al.}(2013)\citenamefont {Her},
  \citenamefont {Beams},\ and\ \citenamefont {Novotny}}]{her2013graphene}%
  \BibitemOpen
  \bibfield  {author} {\bibinfo {author} {\bibfnamefont {Michael}\ \bibnamefont
  {Her}}, \bibinfo {author} {\bibfnamefont {Ryan}\ \bibnamefont {Beams}}, \
  and\ \bibinfo {author} {\bibfnamefont {Lukas}\ \bibnamefont {Novotny}},\
  }\bibfield  {title} {\enquote {\bibinfo {title} {Graphene transfer with
  reduced residue},}\ }\href@noop {} {\bibfield  {journal} {\bibinfo  {journal}
  {Physics Letters A}\ }\textbf {\bibinfo {volume} {377}},\ \bibinfo {pages}
  {1455--1458} (\bibinfo {year} {2013})}\BibitemShut {NoStop}%
\bibitem [{\citenamefont {Lin}\ \emph {et~al.}(2011)\citenamefont {Lin},
  \citenamefont {Lu}, \citenamefont {Yeh}, \citenamefont {Jin}, \citenamefont
  {Suenaga},\ and\ \citenamefont {Chiu}}]{lin2011graphene}%
  \BibitemOpen
  \bibfield  {author} {\bibinfo {author} {\bibfnamefont {Yung-Chang}\
  \bibnamefont {Lin}}, \bibinfo {author} {\bibfnamefont {Chun-Chieh}\
  \bibnamefont {Lu}}, \bibinfo {author} {\bibfnamefont {Chao-Huei}\
  \bibnamefont {Yeh}}, \bibinfo {author} {\bibfnamefont {Chuanhong}\
  \bibnamefont {Jin}}, \bibinfo {author} {\bibfnamefont {Kazu}\ \bibnamefont
  {Suenaga}}, \ and\ \bibinfo {author} {\bibfnamefont {Po-Wen}\ \bibnamefont
  {Chiu}},\ }\bibfield  {title} {\enquote {\bibinfo {title} {Graphene
  annealing: how clean can it be?}}\ }\href@noop {} {\bibfield  {journal}
  {\bibinfo  {journal} {Nano letters}\ }\textbf {\bibinfo {volume} {12}},\
  \bibinfo {pages} {414--419} (\bibinfo {year} {2011})}\BibitemShut {NoStop}%
\bibitem [{\citenamefont {Pirkle}\ \emph {et~al.}(2011)\citenamefont {Pirkle},
  \citenamefont {Chan}, \citenamefont {Venugopal}, \citenamefont {Hinojos},
  \citenamefont {Magnuson}, \citenamefont {McDonnell}, \citenamefont {Colombo},
  \citenamefont {Vogel}, \citenamefont {Ruoff},\ and\ \citenamefont
  {Wallace}}]{pirkle2011effect}%
  \BibitemOpen
  \bibfield  {author} {\bibinfo {author} {\bibfnamefont {A}~\bibnamefont
  {Pirkle}}, \bibinfo {author} {\bibfnamefont {J}~\bibnamefont {Chan}},
  \bibinfo {author} {\bibfnamefont {A}~\bibnamefont {Venugopal}}, \bibinfo
  {author} {\bibfnamefont {D}~\bibnamefont {Hinojos}}, \bibinfo {author}
  {\bibfnamefont {CW}~\bibnamefont {Magnuson}}, \bibinfo {author}
  {\bibfnamefont {S}~\bibnamefont {McDonnell}}, \bibinfo {author}
  {\bibfnamefont {L}~\bibnamefont {Colombo}}, \bibinfo {author} {\bibfnamefont
  {EM}~\bibnamefont {Vogel}}, \bibinfo {author} {\bibfnamefont
  {RS}~\bibnamefont {Ruoff}}, \ and\ \bibinfo {author} {\bibfnamefont
  {RM}~\bibnamefont {Wallace}},\ }\bibfield  {title} {\enquote {\bibinfo
  {title} {The effect of chemical residues on the physical and electrical
  properties of chemical vapor deposited graphene transferred to sio2},}\
  }\href@noop {} {\bibfield  {journal} {\bibinfo  {journal} {Applied Physics
  Letters}\ }\textbf {\bibinfo {volume} {99}},\ \bibinfo {pages} {122108}
  (\bibinfo {year} {2011})}\BibitemShut {NoStop}%
\bibitem [{\citenamefont {Suk}\ \emph {et~al.}(2013)\citenamefont {Suk},
  \citenamefont {Lee}, \citenamefont {Lee}, \citenamefont {Chou}, \citenamefont
  {Piner}, \citenamefont {Hao}, \citenamefont {Akinwande},\ and\ \citenamefont
  {Ruoff}}]{suk2013enhancement}%
  \BibitemOpen
  \bibfield  {author} {\bibinfo {author} {\bibfnamefont {Ji~Won}\ \bibnamefont
  {Suk}}, \bibinfo {author} {\bibfnamefont {Wi~Hyoung}\ \bibnamefont {Lee}},
  \bibinfo {author} {\bibfnamefont {Jongho}\ \bibnamefont {Lee}}, \bibinfo
  {author} {\bibfnamefont {Harry}\ \bibnamefont {Chou}}, \bibinfo {author}
  {\bibfnamefont {Richard~D}\ \bibnamefont {Piner}}, \bibinfo {author}
  {\bibfnamefont {Yufeng}\ \bibnamefont {Hao}}, \bibinfo {author}
  {\bibfnamefont {Deji}\ \bibnamefont {Akinwande}}, \ and\ \bibinfo {author}
  {\bibfnamefont {Rodney~S}\ \bibnamefont {Ruoff}},\ }\bibfield  {title}
  {\enquote {\bibinfo {title} {Enhancement of the electrical properties of
  graphene grown by chemical vapor deposition via controlling the effects of
  polymer residue},}\ }\href@noop {} {\bibfield  {journal} {\bibinfo  {journal}
  {Nano letters}\ }\textbf {\bibinfo {volume} {13}},\ \bibinfo {pages}
  {1462--1467} (\bibinfo {year} {2013})}\BibitemShut {NoStop}%
\bibitem [{\citenamefont {Chan}\ \emph {et~al.}(2012)\citenamefont {Chan},
  \citenamefont {Venugopal}, \citenamefont {Pirkle}, \citenamefont {McDonnell},
  \citenamefont {Hinojos}, \citenamefont {Magnuson}, \citenamefont {Ruoff},
  \citenamefont {Colombo}, \citenamefont {Wallace},\ and\ \citenamefont
  {Vogel}}]{chan2012reducing}%
  \BibitemOpen
  \bibfield  {author} {\bibinfo {author} {\bibfnamefont {Jack}\ \bibnamefont
  {Chan}}, \bibinfo {author} {\bibfnamefont {Archana}\ \bibnamefont
  {Venugopal}}, \bibinfo {author} {\bibfnamefont {Adam}\ \bibnamefont
  {Pirkle}}, \bibinfo {author} {\bibfnamefont {Stephen}\ \bibnamefont
  {McDonnell}}, \bibinfo {author} {\bibfnamefont {David}\ \bibnamefont
  {Hinojos}}, \bibinfo {author} {\bibfnamefont {Carl~W}\ \bibnamefont
  {Magnuson}}, \bibinfo {author} {\bibfnamefont {Rodney~S}\ \bibnamefont
  {Ruoff}}, \bibinfo {author} {\bibfnamefont {Luigi}\ \bibnamefont {Colombo}},
  \bibinfo {author} {\bibfnamefont {Robert~M}\ \bibnamefont {Wallace}}, \ and\
  \bibinfo {author} {\bibfnamefont {Eric~M}\ \bibnamefont {Vogel}},\ }\bibfield
   {title} {\enquote {\bibinfo {title} {Reducing extrinsic performance-limiting
  factors in graphene grown by chemical vapor deposition},}\ }\href@noop {}
  {\bibfield  {journal} {\bibinfo  {journal} {ACS nano}\ }\textbf {\bibinfo
  {volume} {6}},\ \bibinfo {pages} {3224--3229} (\bibinfo {year}
  {2012})}\BibitemShut {NoStop}%
\bibitem [{\citenamefont {Goossens}\ \emph {et~al.}(2012)\citenamefont
  {Goossens}, \citenamefont {Calado}, \citenamefont {Barreiro}, \citenamefont
  {Watanabe}, \citenamefont {Taniguchi},\ and\ \citenamefont
  {Vandersypen}}]{goossens2012mechanical}%
  \BibitemOpen
  \bibfield  {author} {\bibinfo {author} {\bibfnamefont {AM}~\bibnamefont
  {Goossens}}, \bibinfo {author} {\bibfnamefont {VE}~\bibnamefont {Calado}},
  \bibinfo {author} {\bibfnamefont {A}~\bibnamefont {Barreiro}}, \bibinfo
  {author} {\bibfnamefont {K}~\bibnamefont {Watanabe}}, \bibinfo {author}
  {\bibfnamefont {T}~\bibnamefont {Taniguchi}}, \ and\ \bibinfo {author}
  {\bibfnamefont {LMK}\ \bibnamefont {Vandersypen}},\ }\bibfield  {title}
  {\enquote {\bibinfo {title} {Mechanical cleaning of graphene},}\ }\href
  {\doibase 10.1063/1.3685504} {\bibfield  {journal} {\bibinfo  {journal}
  {Applied Physics Letters}\ }\textbf {\bibinfo {volume} {100}},\ \bibinfo
  {pages} {073110} (\bibinfo {year} {2012})}\BibitemShut {NoStop}%
\bibitem [{\citenamefont {Moser}\ \emph {et~al.}(2007)\citenamefont {Moser},
  \citenamefont {Barreiro},\ and\ \citenamefont {Bachtold}}]{moser2007current}%
  \BibitemOpen
  \bibfield  {author} {\bibinfo {author} {\bibfnamefont {Joel}\ \bibnamefont
  {Moser}}, \bibinfo {author} {\bibfnamefont {Amelia}\ \bibnamefont
  {Barreiro}}, \ and\ \bibinfo {author} {\bibfnamefont {Adrian}\ \bibnamefont
  {Bachtold}},\ }\bibfield  {title} {\enquote {\bibinfo {title}
  {Current-induced cleaning of graphene},}\ }\href@noop {} {\bibfield
  {journal} {\bibinfo  {journal} {Applied Physics Letters}\ }\textbf {\bibinfo
  {volume} {91}},\ \bibinfo {pages} {163513} (\bibinfo {year}
  {2007})}\BibitemShut {NoStop}%
\bibitem [{\citenamefont {Pettes}\ \emph {et~al.}(2011)\citenamefont {Pettes},
  \citenamefont {Jo}, \citenamefont {Yao},\ and\ \citenamefont
  {Shi}}]{pettes2011influence}%
  \BibitemOpen
  \bibfield  {author} {\bibinfo {author} {\bibfnamefont {Michael~Thompson}\
  \bibnamefont {Pettes}}, \bibinfo {author} {\bibfnamefont {Insun}\
  \bibnamefont {Jo}}, \bibinfo {author} {\bibfnamefont {Zhen}\ \bibnamefont
  {Yao}}, \ and\ \bibinfo {author} {\bibfnamefont {Li}~\bibnamefont {Shi}},\
  }\bibfield  {title} {\enquote {\bibinfo {title} {Influence of polymeric
  residue on the thermal conductivity of suspended bilayer graphene},}\ }\href
  {\doibase 10.1021/nl104156y} {\bibfield  {journal} {\bibinfo  {journal} {Nano
  Letters}\ }\textbf {\bibinfo {volume} {11}},\ \bibinfo {pages} {1195--1200}
  (\bibinfo {year} {2011})}\BibitemShut {NoStop}%
\bibitem [{\citenamefont {Jo}\ \emph {et~al.}(2015)\citenamefont {Jo},
  \citenamefont {Pettes}, \citenamefont {Lindsay}, \citenamefont {Ou},
  \citenamefont {Weathers}, \citenamefont {Moore}, \citenamefont {Yao},\ and\
  \citenamefont {Shi}}]{jo2015reexamination}%
  \BibitemOpen
  \bibfield  {author} {\bibinfo {author} {\bibfnamefont {Insun}\ \bibnamefont
  {Jo}}, \bibinfo {author} {\bibfnamefont {Michael~T}\ \bibnamefont {Pettes}},
  \bibinfo {author} {\bibfnamefont {Lucas}\ \bibnamefont {Lindsay}}, \bibinfo
  {author} {\bibfnamefont {Eric}\ \bibnamefont {Ou}}, \bibinfo {author}
  {\bibfnamefont {Annie}\ \bibnamefont {Weathers}}, \bibinfo {author}
  {\bibfnamefont {Arden~L}\ \bibnamefont {Moore}}, \bibinfo {author}
  {\bibfnamefont {Zhen}\ \bibnamefont {Yao}}, \ and\ \bibinfo {author}
  {\bibfnamefont {Li}~\bibnamefont {Shi}},\ }\bibfield  {title} {\enquote
  {\bibinfo {title} {Reexamination of basal plane thermal conductivity of
  suspended graphene samples measured by electro-thermal micro-bridge
  methods},}\ }\href {\doibase 10.1063/1.4921519} {\bibfield  {journal}
  {\bibinfo  {journal} {AIP Advances}\ }\textbf {\bibinfo {volume} {5}},\
  \bibinfo {pages} {053206} (\bibinfo {year} {2015})}\BibitemShut {NoStop}%
\bibitem [{\citenamefont {Bunch}\ \emph {et~al.}(2008)\citenamefont {Bunch},
  \citenamefont {Verbridge}, \citenamefont {Alden}, \citenamefont {Van
  Der~Zande}, \citenamefont {Parpia}, \citenamefont {Craighead},\ and\
  \citenamefont {McEuen}}]{bunch2008impermeable}%
  \BibitemOpen
  \bibfield  {author} {\bibinfo {author} {\bibfnamefont {J~Scott}\ \bibnamefont
  {Bunch}}, \bibinfo {author} {\bibfnamefont {Scott~S}\ \bibnamefont
  {Verbridge}}, \bibinfo {author} {\bibfnamefont {Jonathan~S}\ \bibnamefont
  {Alden}}, \bibinfo {author} {\bibfnamefont {Arend~M}\ \bibnamefont {Van
  Der~Zande}}, \bibinfo {author} {\bibfnamefont {Jeevak~M}\ \bibnamefont
  {Parpia}}, \bibinfo {author} {\bibfnamefont {Harold~G}\ \bibnamefont
  {Craighead}}, \ and\ \bibinfo {author} {\bibfnamefont {Paul~L}\ \bibnamefont
  {McEuen}},\ }\bibfield  {title} {\enquote {\bibinfo {title} {Impermeable
  atomic membranes from graphene sheets},}\ }\href {\doibase 10.1021/nl801457b}
  {\bibfield  {journal} {\bibinfo  {journal} {Nano Letters}\ }\textbf {\bibinfo
  {volume} {8}},\ \bibinfo {pages} {2458--2462} (\bibinfo {year}
  {2008})}\BibitemShut {NoStop}%
\bibitem [{\citenamefont {Singh}\ \emph {et~al.}(2010)\citenamefont {Singh},
  \citenamefont {Sengupta}, \citenamefont {Solanki}, \citenamefont {Dhall},
  \citenamefont {Allain}, \citenamefont {Dhara}, \citenamefont {Pant},\ and\
  \citenamefont {Deshmukh}}]{singh2010probing}%
  \BibitemOpen
  \bibfield  {author} {\bibinfo {author} {\bibfnamefont {Vibhor}\ \bibnamefont
  {Singh}}, \bibinfo {author} {\bibfnamefont {Shamashis}\ \bibnamefont
  {Sengupta}}, \bibinfo {author} {\bibfnamefont {Hari~S}\ \bibnamefont
  {Solanki}}, \bibinfo {author} {\bibfnamefont {Rohan}\ \bibnamefont {Dhall}},
  \bibinfo {author} {\bibfnamefont {Adrien}\ \bibnamefont {Allain}}, \bibinfo
  {author} {\bibfnamefont {Sajal}\ \bibnamefont {Dhara}}, \bibinfo {author}
  {\bibfnamefont {Prita}\ \bibnamefont {Pant}}, \ and\ \bibinfo {author}
  {\bibfnamefont {Mandar~M}\ \bibnamefont {Deshmukh}},\ }\bibfield  {title}
  {\enquote {\bibinfo {title} {Probing thermal expansion of graphene and modal
  dispersion at low-temperature using graphene nanoelectromechanical systems
  resonators},}\ }\href {\doibase 10.1088/0957-4484/21/16/165204} {\bibfield
  {journal} {\bibinfo  {journal} {Nanotechnology}\ }\textbf {\bibinfo {volume}
  {21}},\ \bibinfo {pages} {165204} (\bibinfo {year} {2010})}\BibitemShut
  {NoStop}%
\bibitem [{\citenamefont {Barton}\ \emph {et~al.}(2012)\citenamefont {Barton},
  \citenamefont {Storch}, \citenamefont {Adiga}, \citenamefont {Sakakibara},
  \citenamefont {Cipriany}, \citenamefont {Ilic}, \citenamefont {Wang},
  \citenamefont {Ong}, \citenamefont {McEuen}, \citenamefont {Parpia},\ and\
  \citenamefont {Craighead}}]{barton2012photothermal}%
  \BibitemOpen
  \bibfield  {author} {\bibinfo {author} {\bibfnamefont {Robert~A}\
  \bibnamefont {Barton}}, \bibinfo {author} {\bibfnamefont {Isaac~R}\
  \bibnamefont {Storch}}, \bibinfo {author} {\bibfnamefont {Vivekananda~P}\
  \bibnamefont {Adiga}}, \bibinfo {author} {\bibfnamefont {Reyu}\ \bibnamefont
  {Sakakibara}}, \bibinfo {author} {\bibfnamefont {Benjamin~R}\ \bibnamefont
  {Cipriany}}, \bibinfo {author} {\bibfnamefont {B}~\bibnamefont {Ilic}},
  \bibinfo {author} {\bibfnamefont {Si~Ping}\ \bibnamefont {Wang}}, \bibinfo
  {author} {\bibfnamefont {Peijie}\ \bibnamefont {Ong}}, \bibinfo {author}
  {\bibfnamefont {Paul~L}\ \bibnamefont {McEuen}}, \bibinfo {author}
  {\bibfnamefont {Jeevak~M}\ \bibnamefont {Parpia}}, \ and\ \bibinfo {author}
  {\bibfnamefont {Harold~G.}\ \bibnamefont {Craighead}},\ }\bibfield  {title}
  {\enquote {\bibinfo {title} {Photothermal self-oscillation and laser cooling
  of graphene optomechanical systems},}\ }\href {\doibase 10.1021/nl302036x}
  {\bibfield  {journal} {\bibinfo  {journal} {Nano Letters}\ }\textbf {\bibinfo
  {volume} {12}},\ \bibinfo {pages} {4681--4686} (\bibinfo {year}
  {2012})}\BibitemShut {NoStop}%
\bibitem [{\citenamefont {Chen}\ \emph {et~al.}(2009)\citenamefont {Chen},
  \citenamefont {Rosenblatt}, \citenamefont {Bolotin}, \citenamefont {Kalb},
  \citenamefont {Kim}, \citenamefont {Kymissis}, \citenamefont {Stormer},
  \citenamefont {Heinz},\ and\ \citenamefont {Hone}}]{chen2009performance}%
  \BibitemOpen
  \bibfield  {author} {\bibinfo {author} {\bibfnamefont {Changyao}\
  \bibnamefont {Chen}}, \bibinfo {author} {\bibfnamefont {Sami}\ \bibnamefont
  {Rosenblatt}}, \bibinfo {author} {\bibfnamefont {Kirill~I}\ \bibnamefont
  {Bolotin}}, \bibinfo {author} {\bibfnamefont {William}\ \bibnamefont {Kalb}},
  \bibinfo {author} {\bibfnamefont {Philip}\ \bibnamefont {Kim}}, \bibinfo
  {author} {\bibfnamefont {Ioannis}\ \bibnamefont {Kymissis}}, \bibinfo
  {author} {\bibfnamefont {Horst~L}\ \bibnamefont {Stormer}}, \bibinfo {author}
  {\bibfnamefont {Tony~F}\ \bibnamefont {Heinz}}, \ and\ \bibinfo {author}
  {\bibfnamefont {James}\ \bibnamefont {Hone}},\ }\bibfield  {title} {\enquote
  {\bibinfo {title} {Performance of monolayer graphene nanomechanical
  resonators with electrical readout},}\ }\href {\doibase
  10.1038/nnano.2009.267} {\bibfield  {journal} {\bibinfo  {journal} {Nature
  Nanotechnology}\ }\textbf {\bibinfo {volume} {4}},\ \bibinfo {pages}
  {861--867} (\bibinfo {year} {2009})}\BibitemShut {NoStop}%
\bibitem [{\citenamefont {Ferrari}\ \emph {et~al.}(2006)\citenamefont
  {Ferrari}, \citenamefont {Meyer}, \citenamefont {Scardaci}, \citenamefont
  {Casiraghi}, \citenamefont {Lazzeri}, \citenamefont {Mauri}, \citenamefont
  {Piscanec}, \citenamefont {Jiang}, \citenamefont {Novoselov}, \citenamefont
  {Roth},\ and\ \citenamefont {Geim}}]{ferrari2006raman}%
  \BibitemOpen
  \bibfield  {author} {\bibinfo {author} {\bibfnamefont {AC}~\bibnamefont
  {Ferrari}}, \bibinfo {author} {\bibfnamefont {JC}~\bibnamefont {Meyer}},
  \bibinfo {author} {\bibfnamefont {V}~\bibnamefont {Scardaci}}, \bibinfo
  {author} {\bibfnamefont {C}~\bibnamefont {Casiraghi}}, \bibinfo {author}
  {\bibfnamefont {Michele}\ \bibnamefont {Lazzeri}}, \bibinfo {author}
  {\bibfnamefont {Francesco}\ \bibnamefont {Mauri}}, \bibinfo {author}
  {\bibfnamefont {S}~\bibnamefont {Piscanec}}, \bibinfo {author} {\bibfnamefont
  {Da}~\bibnamefont {Jiang}}, \bibinfo {author} {\bibfnamefont
  {KS}~\bibnamefont {Novoselov}}, \bibinfo {author} {\bibfnamefont
  {S}~\bibnamefont {Roth}}, \ and\ \bibinfo {author} {\bibfnamefont
  {AK}~\bibnamefont {Geim}},\ }\bibfield  {title} {\enquote {\bibinfo {title}
  {Raman spectrum of graphene and graphene layers},}\ }\href@noop {} {\bibfield
   {journal} {\bibinfo  {journal} {Physical Review Letters}\ }\textbf {\bibinfo
  {volume} {97}},\ \bibinfo {pages} {187401} (\bibinfo {year}
  {2006})}\BibitemShut {NoStop}%
\bibitem [{\citenamefont {Blake}\ \emph {et~al.}(2007)\citenamefont {Blake},
  \citenamefont {Hill}, \citenamefont {Castro~Neto}, \citenamefont {Novoselov},
  \citenamefont {Jiang}, \citenamefont {Yang}, \citenamefont {Booth},\ and\
  \citenamefont {Geim}}]{blake2007making}%
  \BibitemOpen
  \bibfield  {author} {\bibinfo {author} {\bibfnamefont {P}~\bibnamefont
  {Blake}}, \bibinfo {author} {\bibfnamefont {E~W}\ \bibnamefont {Hill}},
  \bibinfo {author} {\bibfnamefont {AH}~\bibnamefont {Castro~Neto}}, \bibinfo
  {author} {\bibfnamefont {K~S}\ \bibnamefont {Novoselov}}, \bibinfo {author}
  {\bibfnamefont {D}~\bibnamefont {Jiang}}, \bibinfo {author} {\bibfnamefont
  {R}~\bibnamefont {Yang}}, \bibinfo {author} {\bibfnamefont {T~J}\
  \bibnamefont {Booth}}, \ and\ \bibinfo {author} {\bibfnamefont
  {AK}~\bibnamefont {Geim}},\ }\bibfield  {title} {\enquote {\bibinfo {title}
  {Making graphene visible},}\ }\href {\doibase 10.1063/1.2768624} {\bibfield
  {journal} {\bibinfo  {journal} {Applied Physics Letters}\ }\textbf {\bibinfo
  {volume} {91}},\ \bibinfo {pages} {063124} (\bibinfo {year}
  {2007})}\BibitemShut {NoStop}%
\bibitem [{\citenamefont {Nemes-Incze}\ \emph {et~al.}(2008)\citenamefont
  {Nemes-Incze}, \citenamefont {Osv{\'a}th}, \citenamefont {Kamar{\'a}s},\ and\
  \citenamefont {Bir{\'o}}}]{nemes2008anomalies}%
  \BibitemOpen
  \bibfield  {author} {\bibinfo {author} {\bibfnamefont {P}~\bibnamefont
  {Nemes-Incze}}, \bibinfo {author} {\bibfnamefont {Z}~\bibnamefont
  {Osv{\'a}th}}, \bibinfo {author} {\bibfnamefont {K}~\bibnamefont
  {Kamar{\'a}s}}, \ and\ \bibinfo {author} {\bibfnamefont {LP}~\bibnamefont
  {Bir{\'o}}},\ }\bibfield  {title} {\enquote {\bibinfo {title} {Anomalies in
  thickness measurements of graphene and few layer graphite crystals by tapping
  mode atomic force microscopy},}\ }\href@noop {} {\bibfield  {journal}
  {\bibinfo  {journal} {Carbon}\ }\textbf {\bibinfo {volume} {46}},\ \bibinfo
  {pages} {1435--1442} (\bibinfo {year} {2008})}\BibitemShut {NoStop}%
\bibitem [{\citenamefont {Nicholl}\ \emph {et~al.}(2015)\citenamefont
  {Nicholl}, \citenamefont {Conley}, \citenamefont {Lavrik}, \citenamefont
  {Vlassiouk}, \citenamefont {Puzyrev}, \citenamefont {Sreenivas},
  \citenamefont {Pantelides},\ and\ \citenamefont
  {Bolotin}}]{nicholl2015effect}%
  \BibitemOpen
  \bibfield  {author} {\bibinfo {author} {\bibfnamefont {Ryan~JT}\ \bibnamefont
  {Nicholl}}, \bibinfo {author} {\bibfnamefont {Hiram~J}\ \bibnamefont
  {Conley}}, \bibinfo {author} {\bibfnamefont {Nickolay~V}\ \bibnamefont
  {Lavrik}}, \bibinfo {author} {\bibfnamefont {Ivan}\ \bibnamefont
  {Vlassiouk}}, \bibinfo {author} {\bibfnamefont {Yevgeniy~S}\ \bibnamefont
  {Puzyrev}}, \bibinfo {author} {\bibfnamefont {Vijayashree~Parsi}\
  \bibnamefont {Sreenivas}}, \bibinfo {author} {\bibfnamefont {Sokrates~T}\
  \bibnamefont {Pantelides}}, \ and\ \bibinfo {author} {\bibfnamefont
  {Kirill~I}\ \bibnamefont {Bolotin}},\ }\bibfield  {title} {\enquote {\bibinfo
  {title} {The effect of intrinsic crumpling on the mechanics of free-standing
  graphene},}\ }\href@noop {} {\bibfield  {journal} {\bibinfo  {journal}
  {Nature communications}\ }\textbf {\bibinfo {volume} {6}},\ \bibinfo {pages}
  {8789} (\bibinfo {year} {2015})}\BibitemShut {NoStop}%
\bibitem [{\citenamefont {Lee}\ \emph {et~al.}(2012)\citenamefont {Lee},
  \citenamefont {Yoon},\ and\ \citenamefont {Cheong}}]{lee2012estimation}%
  \BibitemOpen
  \bibfield  {author} {\bibinfo {author} {\bibfnamefont {Jae-Ung}\ \bibnamefont
  {Lee}}, \bibinfo {author} {\bibfnamefont {Duhee}\ \bibnamefont {Yoon}}, \
  and\ \bibinfo {author} {\bibfnamefont {Hyeonsik}\ \bibnamefont {Cheong}},\
  }\bibfield  {title} {\enquote {\bibinfo {title} {Estimation of {Y}oung’s
  modulus of graphene by {Raman} spectroscopy},}\ }\href@noop {} {\bibfield
  {journal} {\bibinfo  {journal} {Nano Letters}\ }\textbf {\bibinfo {volume}
  {12}},\ \bibinfo {pages} {4444--4448} (\bibinfo {year} {2012})}\BibitemShut
  {NoStop}%
\bibitem [{\citenamefont {Ruiz-Vargas}\ \emph {et~al.}(2011)\citenamefont
  {Ruiz-Vargas}, \citenamefont {Zhuang}, \citenamefont {Huang}, \citenamefont
  {Van Der~Zande}, \citenamefont {Garg}, \citenamefont {McEuen}, \citenamefont
  {Muller}, \citenamefont {Hennig},\ and\ \citenamefont
  {Park}}]{ruiz2011softened}%
  \BibitemOpen
  \bibfield  {author} {\bibinfo {author} {\bibfnamefont {Carlos~S}\
  \bibnamefont {Ruiz-Vargas}}, \bibinfo {author} {\bibfnamefont {Houlong~L}\
  \bibnamefont {Zhuang}}, \bibinfo {author} {\bibfnamefont {Pinshane~Y}\
  \bibnamefont {Huang}}, \bibinfo {author} {\bibfnamefont {Arend~M}\
  \bibnamefont {Van Der~Zande}}, \bibinfo {author} {\bibfnamefont {Shivank}\
  \bibnamefont {Garg}}, \bibinfo {author} {\bibfnamefont {Paul~L}\ \bibnamefont
  {McEuen}}, \bibinfo {author} {\bibfnamefont {David~A}\ \bibnamefont
  {Muller}}, \bibinfo {author} {\bibfnamefont {Richard~G}\ \bibnamefont
  {Hennig}}, \ and\ \bibinfo {author} {\bibfnamefont {Jiwoong}\ \bibnamefont
  {Park}},\ }\bibfield  {title} {\enquote {\bibinfo {title} {Softened elastic
  response and unzipping in chemical vapor deposition graphene membranes},}\
  }\href@noop {} {\bibfield  {journal} {\bibinfo  {journal} {Nano letters}\
  }\textbf {\bibinfo {volume} {11}},\ \bibinfo {pages} {2259--2263} (\bibinfo
  {year} {2011})}\BibitemShut {NoStop}%
\bibitem [{\citenamefont {Davidovikj}\ \emph {et~al.}(2016)\citenamefont
  {Davidovikj}, \citenamefont {Slim}, \citenamefont {Cartamil-Bueno},
  \citenamefont {van~der Zant}, \citenamefont {Steeneken},\ and\ \citenamefont
  {Venstra}}]{davidovikj2016visualizing}%
  \BibitemOpen
  \bibfield  {author} {\bibinfo {author} {\bibfnamefont {Dejan}\ \bibnamefont
  {Davidovikj}}, \bibinfo {author} {\bibfnamefont {Jesse~J}\ \bibnamefont
  {Slim}}, \bibinfo {author} {\bibfnamefont {Santiago~J}\ \bibnamefont
  {Cartamil-Bueno}}, \bibinfo {author} {\bibfnamefont {Herre S~J}\ \bibnamefont
  {van~der Zant}}, \bibinfo {author} {\bibfnamefont {Peter~G}\ \bibnamefont
  {Steeneken}}, \ and\ \bibinfo {author} {\bibfnamefont {Warner~J}\
  \bibnamefont {Venstra}},\ }\bibfield  {title} {\enquote {\bibinfo {title}
  {Visualizing the motion of graphene nanodrums},}\ }\href {\doibase
  10.1021/acs.nanolett.6b00477} {\bibfield  {journal} {\bibinfo  {journal}
  {Nano Letters}\ }\textbf {\bibinfo {volume} {16}},\ \bibinfo {pages}
  {2768--2773} (\bibinfo {year} {2016})}\BibitemShut {NoStop}%
\bibitem [{\citenamefont {Wong}\ \emph {et~al.}(2010)\citenamefont {Wong},
  \citenamefont {Annamalai}, \citenamefont {Wang},\ and\ \citenamefont
  {Palaniapan}}]{wong2010characterization}%
  \BibitemOpen
  \bibfield  {author} {\bibinfo {author} {\bibfnamefont {CL}~\bibnamefont
  {Wong}}, \bibinfo {author} {\bibfnamefont {M}~\bibnamefont {Annamalai}},
  \bibinfo {author} {\bibfnamefont {ZQ}~\bibnamefont {Wang}}, \ and\ \bibinfo
  {author} {\bibfnamefont {M}~\bibnamefont {Palaniapan}},\ }\bibfield  {title}
  {\enquote {\bibinfo {title} {Characterization of nanomechanical graphene drum
  structures},}\ }\href {\doibase 10.1088/0960-1317/20/11/115029} {\bibfield
  {journal} {\bibinfo  {journal} {Journal of Micromechanics and
  Microengineering}\ }\textbf {\bibinfo {volume} {20}},\ \bibinfo {pages}
  {115029} (\bibinfo {year} {2010})}\BibitemShut {NoStop}%
\bibitem [{\citenamefont {Sauerbrey}(1959)}]{Sauerbrey1959}%
  \BibitemOpen
  \bibfield  {author} {\bibinfo {author} {\bibfnamefont {G{\"u}nter}\
  \bibnamefont {Sauerbrey}},\ }\bibfield  {title} {\enquote {\bibinfo {title}
  {Verwendung von schwingquarzen zur w{\"a}gung d{\"u}nner schichten und zur
  mikrow{\"a}gung},}\ }\href {\doibase 10.1007/BF01337937} {\bibfield
  {journal} {\bibinfo  {journal} {Zeitschrift f{\"u}r Physik}\ }\textbf
  {\bibinfo {volume} {155}},\ \bibinfo {pages} {206--222} (\bibinfo {year}
  {1959})}\BibitemShut {NoStop}%
\bibitem [{\citenamefont {O’sullivan}\ and\ \citenamefont
  {Guilbault}(1999)}]{o1999commercial}%
  \BibitemOpen
  \bibfield  {author} {\bibinfo {author} {\bibfnamefont {CK}~\bibnamefont
  {O’sullivan}}\ and\ \bibinfo {author} {\bibfnamefont {GG}~\bibnamefont
  {Guilbault}},\ }\bibfield  {title} {\enquote {\bibinfo {title} {Commercial
  quartz crystal microbalances--theory and applications},}\ }\href@noop {}
  {\bibfield  {journal} {\bibinfo  {journal} {Biosensors and bioelectronics}\
  }\textbf {\bibinfo {volume} {14}},\ \bibinfo {pages} {663--670} (\bibinfo
  {year} {1999})}\BibitemShut {NoStop}%
\bibitem [{\citenamefont {Buttry}\ and\ \citenamefont
  {Ward}(1992)}]{buttry1992measurement}%
  \BibitemOpen
  \bibfield  {author} {\bibinfo {author} {\bibfnamefont {Daniel~A}\
  \bibnamefont {Buttry}}\ and\ \bibinfo {author} {\bibfnamefont {Michael~D}\
  \bibnamefont {Ward}},\ }\bibfield  {title} {\enquote {\bibinfo {title}
  {Measurement of interfacial processes at electrode surfaces with the
  electrochemical quartz crystal microbalance},}\ }\href@noop {} {\bibfield
  {journal} {\bibinfo  {journal} {Chemical Reviews}\ }\textbf {\bibinfo
  {volume} {92}},\ \bibinfo {pages} {1355--1379} (\bibinfo {year}
  {1992})}\BibitemShut {NoStop}%
\bibitem [{\citenamefont {Li}\ \emph {et~al.}(2015)\citenamefont {Li},
  \citenamefont {Song}, \citenamefont {Besenbacher},\ and\ \citenamefont
  {Dong}}]{doi:10.1021/ar500306w}%
  \BibitemOpen
  \bibfield  {author} {\bibinfo {author} {\bibfnamefont {Qiang}\ \bibnamefont
  {Li}}, \bibinfo {author} {\bibfnamefont {Jie}\ \bibnamefont {Song}}, \bibinfo
  {author} {\bibfnamefont {Flemming}\ \bibnamefont {Besenbacher}}, \ and\
  \bibinfo {author} {\bibfnamefont {Mingdong}\ \bibnamefont {Dong}},\
  }\bibfield  {title} {\enquote {\bibinfo {title} {Two-dimensional material
  confined water},}\ }\href {\doibase 10.1021/ar500306w} {\bibfield  {journal}
  {\bibinfo  {journal} {Accounts of Chemical Research}\ }\textbf {\bibinfo
  {volume} {48}},\ \bibinfo {pages} {119--127} (\bibinfo {year}
  {2015})}\BibitemShut {NoStop}%
\bibitem [{\citenamefont {Davidovikj}\ \emph {et~al.}(2018)\citenamefont
  {Davidovikj}, \citenamefont {Poot}, \citenamefont {Cartamil-Bueno},
  \citenamefont {van~der Zant},\ and\ \citenamefont
  {Steeneken}}]{davidovikj2017thermal}%
  \BibitemOpen
  \bibfield  {author} {\bibinfo {author} {\bibfnamefont {Dejan}\ \bibnamefont
  {Davidovikj}}, \bibinfo {author} {\bibfnamefont {Menno}\ \bibnamefont
  {Poot}}, \bibinfo {author} {\bibfnamefont {Santiago~J}\ \bibnamefont
  {Cartamil-Bueno}}, \bibinfo {author} {\bibfnamefont {Herre~SJ}\ \bibnamefont
  {van~der Zant}}, \ and\ \bibinfo {author} {\bibfnamefont {Peter~G}\
  \bibnamefont {Steeneken}},\ }\bibfield  {title} {\enquote {\bibinfo {title}
  {On-chip heaters for tension tuning of graphene nanodrums},}\ }\href
  {\doibase 10.1021/acs.nanolett.7b05358} {\bibfield  {journal} {\bibinfo
  {journal} {Nano Letters}\ }\textbf {\bibinfo {volume} {18}},\ \bibinfo
  {pages} {2852--2858} (\bibinfo {year} {2018})}\BibitemShut {NoStop}%
\bibitem [{\citenamefont {Ye}\ \emph {et~al.}(2017)\citenamefont {Ye},
  \citenamefont {Lee},\ and\ \citenamefont {Feng}}]{ye2017very}%
  \BibitemOpen
  \bibfield  {author} {\bibinfo {author} {\bibfnamefont {Fan}\ \bibnamefont
  {Ye}}, \bibinfo {author} {\bibfnamefont {Jaesung}\ \bibnamefont {Lee}}, \
  and\ \bibinfo {author} {\bibfnamefont {Philip X-L}\ \bibnamefont {Feng}},\
  }\bibfield  {title} {\enquote {\bibinfo {title} {Very-wide electrothermal
  tuning of graphene nanoelectromechanical resonators},}\ }in\ \href {\doibase
  10.1109/MEMSYS.2017.7863341} {\emph {\bibinfo {booktitle} {Micro Electro
  Mechanical Systems (MEMS), 2017 IEEE 30th International Conference on}}}\
  (\bibinfo {organization} {IEEE},\ \bibinfo {year} {2017})\ pp.\ \bibinfo
  {pages} {68--71}\BibitemShut {NoStop}%
\bibitem [{\citenamefont {Ghosh}\ \emph {et~al.}(2008)\citenamefont {Ghosh},
  \citenamefont {Calizo}, \citenamefont {Teweldebrhan}, \citenamefont
  {Pokatilov}, \citenamefont {Nika}, \citenamefont {Balandin}, \citenamefont
  {Bao}, \citenamefont {Miao},\ and\ \citenamefont {Lau}}]{ghosh2008extremely}%
  \BibitemOpen
  \bibfield  {author} {\bibinfo {author} {\bibfnamefont {S}~\bibnamefont
  {Ghosh}}, \bibinfo {author} {\bibfnamefont {I}~\bibnamefont {Calizo}},
  \bibinfo {author} {\bibfnamefont {D}~\bibnamefont {Teweldebrhan}}, \bibinfo
  {author} {\bibfnamefont {EP}~\bibnamefont {Pokatilov}}, \bibinfo {author}
  {\bibfnamefont {DL}~\bibnamefont {Nika}}, \bibinfo {author} {\bibfnamefont
  {AA}~\bibnamefont {Balandin}}, \bibinfo {author} {\bibfnamefont
  {W}~\bibnamefont {Bao}}, \bibinfo {author} {\bibfnamefont {F}~\bibnamefont
  {Miao}}, \ and\ \bibinfo {author} {\bibfnamefont {C~Ning}\ \bibnamefont
  {Lau}},\ }\bibfield  {title} {\enquote {\bibinfo {title} {Extremely high
  thermal conductivity of graphene: Prospects for thermal management
  applications in nanoelectronic circuits},}\ }\href {\doibase
  10.1063/1.2907977} {\bibfield  {journal} {\bibinfo  {journal} {Applied
  Physics Letters}\ }\textbf {\bibinfo {volume} {92}},\ \bibinfo {pages}
  {151911} (\bibinfo {year} {2008})}\BibitemShut {NoStop}%
\bibitem [{\citenamefont {Nicholl}\ \emph {et~al.}(2017)\citenamefont
  {Nicholl}, \citenamefont {Lavrik}, \citenamefont {Vlassiouk}, \citenamefont
  {Srijanto},\ and\ \citenamefont {Bolotin}}]{PhysRevLett.118.266101}%
  \BibitemOpen
  \bibfield  {author} {\bibinfo {author} {\bibfnamefont {Ryan J.~T.}\
  \bibnamefont {Nicholl}}, \bibinfo {author} {\bibfnamefont {Nickolay~V.}\
  \bibnamefont {Lavrik}}, \bibinfo {author} {\bibfnamefont {Ivan}\ \bibnamefont
  {Vlassiouk}}, \bibinfo {author} {\bibfnamefont {Bernadeta~R.}\ \bibnamefont
  {Srijanto}}, \ and\ \bibinfo {author} {\bibfnamefont {Kirill~I.}\
  \bibnamefont {Bolotin}},\ }\bibfield  {title} {\enquote {\bibinfo {title}
  {Hidden area and mechanical nonlinearities in freestanding graphene},}\
  }\href {\doibase 10.1103/PhysRevLett.118.266101} {\bibfield  {journal}
  {\bibinfo  {journal} {Physical Review Letters}\ }\textbf {\bibinfo {volume}
  {118}},\ \bibinfo {pages} {266101} (\bibinfo {year} {2017})}\BibitemShut
  {NoStop}%
\bibitem [{\citenamefont {Cartamil-Bueno}\ \emph {et~al.}(2017)\citenamefont
  {Cartamil-Bueno}, \citenamefont {Cavalieri}, \citenamefont {Wang},
  \citenamefont {Houri}, \citenamefont {Hofmann},\ and\ \citenamefont {van~der
  Zant}}]{cartamil2017mechanical}%
  \BibitemOpen
  \bibfield  {author} {\bibinfo {author} {\bibfnamefont {Santiago~J}\
  \bibnamefont {Cartamil-Bueno}}, \bibinfo {author} {\bibfnamefont {Matteo}\
  \bibnamefont {Cavalieri}}, \bibinfo {author} {\bibfnamefont {Ruizhi}\
  \bibnamefont {Wang}}, \bibinfo {author} {\bibfnamefont {Samer}\ \bibnamefont
  {Houri}}, \bibinfo {author} {\bibfnamefont {Stephan}\ \bibnamefont
  {Hofmann}}, \ and\ \bibinfo {author} {\bibfnamefont {Herre~SJ}\ \bibnamefont
  {van~der Zant}},\ }\bibfield  {title} {\enquote {\bibinfo {title} {Mechanical
  characterization and cleaning of cvd single-layer h-bn resonators},}\
  }\href@noop {} {\bibfield  {journal} {\bibinfo  {journal} {npj 2D Materials
  and Applications}\ }\textbf {\bibinfo {volume} {1}},\ \bibinfo {pages} {16}
  (\bibinfo {year} {2017})}\BibitemShut {NoStop}%
\bibitem [{\citenamefont {Thanner}\ \emph {et~al.}(2002)\citenamefont
  {Thanner}, \citenamefont {Krempl}, \citenamefont {Walln{\"o}fer},\ and\
  \citenamefont {Worsch}}]{thanner2002gapo4}%
  \BibitemOpen
  \bibfield  {author} {\bibinfo {author} {\bibfnamefont {H}~\bibnamefont
  {Thanner}}, \bibinfo {author} {\bibfnamefont {PW}~\bibnamefont {Krempl}},
  \bibinfo {author} {\bibfnamefont {W}~\bibnamefont {Walln{\"o}fer}}, \ and\
  \bibinfo {author} {\bibfnamefont {PM}~\bibnamefont {Worsch}},\ }\bibfield
  {title} {\enquote {\bibinfo {title} {Gapo$_4$ high temperature crystal
  microbalance with zero temperature coefficient},}\ }\href@noop {} {\bibfield
  {journal} {\bibinfo  {journal} {Vacuum}\ }\textbf {\bibinfo {volume} {67}},\
  \bibinfo {pages} {687--691} (\bibinfo {year} {2002})}\BibitemShut {NoStop}%
\bibitem [{\citenamefont {Thanner}\ \emph {et~al.}(2003)\citenamefont
  {Thanner}, \citenamefont {Krempl}, \citenamefont {Selic}, \citenamefont
  {Walln{\"o}fer},\ and\ \citenamefont {Worsch}}]{Thanner2003}%
  \BibitemOpen
  \bibfield  {author} {\bibinfo {author} {\bibfnamefont {H.}~\bibnamefont
  {Thanner}}, \bibinfo {author} {\bibfnamefont {P.~W.}\ \bibnamefont {Krempl}},
  \bibinfo {author} {\bibfnamefont {R.}~\bibnamefont {Selic}}, \bibinfo
  {author} {\bibfnamefont {W.}~\bibnamefont {Walln{\"o}fer}}, \ and\ \bibinfo
  {author} {\bibfnamefont {P.~M.}\ \bibnamefont {Worsch}},\ }\bibfield  {title}
  {\enquote {\bibinfo {title} {Ga{PO}$_4$ high temperature crystal microbalance
  demonstration up to 720{$^{\circ}$C}},}\ }\href {\doibase
  10.1023/A:1022249713714} {\bibfield  {journal} {\bibinfo  {journal} {Journal
  of Thermal Analysis and Calorimetry}\ }\textbf {\bibinfo {volume} {71}},\
  \bibinfo {pages} {53--59} (\bibinfo {year} {2003})}\BibitemShut {NoStop}%
\bibitem [{\citenamefont {Liang}\ \emph {et~al.}(2011)\citenamefont {Liang},
  \citenamefont {Sperling}, \citenamefont {Calizo}, \citenamefont {Cheng},
  \citenamefont {Hacker}, \citenamefont {Zhang}, \citenamefont {Obeng},
  \citenamefont {Yan}, \citenamefont {Peng}, \citenamefont {Li} \emph
  {et~al.}}]{liang2011toward}%
  \BibitemOpen
  \bibfield  {author} {\bibinfo {author} {\bibfnamefont {Xuelei}\ \bibnamefont
  {Liang}}, \bibinfo {author} {\bibfnamefont {Brent~A}\ \bibnamefont
  {Sperling}}, \bibinfo {author} {\bibfnamefont {Irene}\ \bibnamefont
  {Calizo}}, \bibinfo {author} {\bibfnamefont {Guangjun}\ \bibnamefont
  {Cheng}}, \bibinfo {author} {\bibfnamefont {Christina~Ann}\ \bibnamefont
  {Hacker}}, \bibinfo {author} {\bibfnamefont {Qin}\ \bibnamefont {Zhang}},
  \bibinfo {author} {\bibfnamefont {Yaw}\ \bibnamefont {Obeng}}, \bibinfo
  {author} {\bibfnamefont {Kai}\ \bibnamefont {Yan}}, \bibinfo {author}
  {\bibfnamefont {Hailin}\ \bibnamefont {Peng}}, \bibinfo {author}
  {\bibfnamefont {Qiliang}\ \bibnamefont {Li}},  \emph {et~al.},\ }\bibfield
  {title} {\enquote {\bibinfo {title} {Toward clean and crackless transfer of
  graphene},}\ }\href@noop {} {\bibfield  {journal} {\bibinfo  {journal} {ACS
  nano}\ }\textbf {\bibinfo {volume} {5}},\ \bibinfo {pages} {9144--9153}
  (\bibinfo {year} {2011})}\BibitemShut {NoStop}%
\bibitem [{\citenamefont {Lee}\ \emph {et~al.}(2015)\citenamefont {Lee},
  \citenamefont {Lee}, \citenamefont {Kang}, \citenamefont {Cho}, \citenamefont
  {Lee}, \citenamefont {Jung},\ and\ \citenamefont {Lee}}]{LEE2015286}%
  \BibitemOpen
  \bibfield  {author} {\bibinfo {author} {\bibfnamefont {Sangchul}\
  \bibnamefont {Lee}}, \bibinfo {author} {\bibfnamefont {Sang~Kyung}\
  \bibnamefont {Lee}}, \bibinfo {author} {\bibfnamefont {Chang~Goo}\
  \bibnamefont {Kang}}, \bibinfo {author} {\bibfnamefont {Chunhum}\
  \bibnamefont {Cho}}, \bibinfo {author} {\bibfnamefont {Young~Gon}\
  \bibnamefont {Lee}}, \bibinfo {author} {\bibfnamefont {Ukjin}\ \bibnamefont
  {Jung}}, \ and\ \bibinfo {author} {\bibfnamefont {Byoung~Hun}\ \bibnamefont
  {Lee}},\ }\bibfield  {title} {\enquote {\bibinfo {title} {Graphene transfer
  in vacuum yielding a high quality graphene},}\ }\href {\doibase
  https://doi.org/10.1016/j.carbon.2015.05.038} {\bibfield  {journal} {\bibinfo
   {journal} {Carbon}\ }\textbf {\bibinfo {volume} {93}},\ \bibinfo {pages}
  {286 -- 294} (\bibinfo {year} {2015})}\BibitemShut {NoStop}%
\bibitem [{\citenamefont {Wu}\ \emph {et~al.}(2018)\citenamefont {Wu},
  \citenamefont {Lin}, \citenamefont {Cong}, \citenamefont {Liu},\ and\
  \citenamefont {Tan}}]{C6CS00915H}%
  \BibitemOpen
  \bibfield  {author} {\bibinfo {author} {\bibfnamefont {Jiang-Bin}\
  \bibnamefont {Wu}}, \bibinfo {author} {\bibfnamefont {Miao-Ling}\
  \bibnamefont {Lin}}, \bibinfo {author} {\bibfnamefont {Xin}\ \bibnamefont
  {Cong}}, \bibinfo {author} {\bibfnamefont {He-Nan}\ \bibnamefont {Liu}}, \
  and\ \bibinfo {author} {\bibfnamefont {Ping-Heng}\ \bibnamefont {Tan}},\
  }\bibfield  {title} {\enquote {\bibinfo {title} {Raman spectroscopy of
  graphene-based materials and its applications in related devices},}\
  }\href@noop {} {\bibfield  {journal} {\bibinfo  {journal} {Chem. Soc. Rev.}\
  }\textbf {\bibinfo {volume} {47}},\ \bibinfo {pages} {1822--1873} (\bibinfo
  {year} {2018})}\BibitemShut {NoStop}%
\end{thebibliography}%

\end{document}